\documentclass[usegraphicx,usenatbib]{mn2e}
\usepackage{aas_macros,graphicx,times,multirow}
\usepackage{amssymb}

\title[X-ray and UV correlation in the quiescent emission of Cen X-4, evidence of accretion and reprocessing]
{Daily, multiwavelength {\it Swift} monitoring of the neutron star low-mass X-ray binary Cen X-4: evidence for accretion and reprocessing during quiescence}
\author[F.~Bernardini et al.]                                                    
{F.~Bernardini$^{1,2}$\thanks{E-mail: bernardini@wayne.edu}, 
E.~M.~Cackett$^{1}$,
E.~F.~Brown$^3$,
C.~D'Angelo$^{4}$,
N.~Degenaar$^{5,6}$,
J.~M.~Miller$^5$,
\newauthor
M.~Reynolds$^5$,
R.~Wijnands$^4$\\
$^1$ Department of Physics \& Astronomy, Wayne State University, 666 W. Hancock St., Detroit, MI 48201, USA\\
$^2$ INAF, Osservatorio Astronomico di Capodimonte, Salita Moiariello 16, 80131 Napoli, Italia\\
$^3$ Department of Physics \& Astronomy, National Superconducting Cyclotron Laboratory, and the Joint Institute for Nuclear Astrophysics,\\ Michigan State University, East Lansing, MI 48824, USA\\
$^4$ Instituut Anton Pannekoek, University of Amsterdam, Amsterdam 1098 XH, The Netherlands \\
$^5$ Department of Astronomy, University of Michigan, 500 Church St, Ann Arbor, MI 48109-1042, USA\\
$^6$ Hubble fellow\\}
\date{}
\pagerange{\pageref{firstpage}--\pageref{lastpage}}
\def\Sw{{\em Swift}}
\def\XMM{{\em XMM-Newton}}

\def\ergscm{\rm erg\,cm$^{-2}$\,s$^{-1}$}

\newcommand\beq{\begin{equation}}
\newcommand\eeq{\end{equation}}

\begin{document}
\label{firstpage}
\maketitle
\begin{abstract}
We conducted the first long-term (60 days), multiwavelength (optical, ultraviolet, and X-ray) simultaneous monitoring of Cen X-4 with daily \Sw\ observations from June to August 2012, with the goal of understanding variability in the low mass X-ray binary Cen X-4 during quiescence. We found Cen X-4 to be highly variable in all energy bands on timescales from days to months, with the strongest quiescent variability a factor of 22 drop in the X-ray count rate in only 4 days. The X-ray, UV and optical (V band) emission are correlated on timescales down to less than 110 s. The shape of the correlation is a power law with index $\gamma$ about 0.2--0.6. The X-ray spectrum is well fitted by a hydrogen NS atmosphere ($kT=59-80$ eV) and a power law (with spectral index $\Gamma=1.4-2.0$), with the spectral shape remaining constant as the flux varies. Both components vary in tandem, with each responsible for about 50\% of the total X-ray flux, implying that they are physically linked. We conclude that the X-rays are likely generated by matter accreting down to the NS surface. Moreover, based on the short timescale of the correlation, we also unambiguously demonstrate that the UV emission can not be due to either thermal emission from the stream impact point, or a standard optically thick, geometrically thin disc. The spectral energy distribution shows a small UV emitting region, too hot to arise from the accretion disk, that we identified as a hot spot on the companion star. Therefore, the UV emission is most likely produced by reprocessing from the companion star, indeed the vertical size of the disc is small and can only reprocess a marginal fraction of the X-ray emission. We also found the accretion disc in quiescence to likely be UV faint, with a minimal contribution to the whole UV flux.  

\end{abstract}

\section{Introduction}

In low mass X-ray binaries (LMXBs), a compact object (a neutron star, NS, or a black hole, BH), is accreting matter from a companion star, a main sequence dwarf, or a white dwarf, or a sub-giant with mass smaller (or much smaller) than that of the Sun. Some LMXBs are persistently bright X-ray sources, because they are accreting matter at high rates from their companion star. Their X-ray emission below 10 keV is $L_{X}\sim10^{37-38}$ erg/s (tens of percent of the Eddington rate), some are persistently not so bright, $L_{X}=10^{35}-10^{36}$ erg/s, others are instead persistently very faint, $L_{X}\sim10^{34-35}$ erg/s, \citep[see e.g.][and references therein]{armas13}. However, some of them are transient and alternate between long periods of quiescence (lasting up to decades) during which the X-ray emission is faint, generally in the range $L_{X}=10^{32}-10^{33}$ erg/s, to sporadic periods of intense emission (outbursts, lasting weeks-months or even years), where the X-ray luminosity strongly increases (up to 6 orders of magnitude at the peak) becoming similar to the persistently bright LMXBs. The physics of LMXBs during their outburst state is thought to be well established. The general properties of the quiescence to outburst cycle is explained by the disc instability model \citep[DIM, e.g.][]{cannizzo93,laso}.  
During outburst, the compact object accretes matter from an accretion disc which has been interpreted has an optically thick and geometrically thin disc \citep{shakura73} extending close to its surface.  On the contrary the mechanism powering the optical, UV and X-ray emission in quiescence is still debated. The emission could be powered by residual accretion, however, the physics of accretion at low Eddington luminosity rates is far from being understood. Several models have been proposed to explain the difference in the emission from an LMXB containing a NS and a BH. However, so far, a unifying scenario which can make a clear prediction, systematically matching the spectral energy distribution (SED), from optical up to the X-ray emission, of both NS and BH quiescent LMXB, is still missing.
 
\cite{brown98} pointed out that the X-ray luminosity of NS LMXB in quiescence should be higher than for BH (deep crustal heating model). Indeed, during an outburst, nuclear reactions are thought to be activated in the upper layers of the NS crust, efficiently transferring energy to the NS core that consequently heats up. Once the crust has thermally relaxed, the energy stored in the core is thermally released on longer timescales, at a sufficient rate to power the quiescent X-ray emission.  This model is best applied to the low quiescent luminosity systems which undergo frequent outbursts.  In this model flux variation (up to a factor of 2--3) only arises between different quiescent epochs due to changes in the composition of the envelope caused by nuclear burning during outburst \citep{brown02}.
Immediately after an outburst ends the crust (which has been heated during the outburst) is expected to thermally relax \citep{rutledge02}, and such crustal cooling has now ben observed in several sources \citep[e.g.][and more reference therein]{cackett08,cackett10_ks1731, cackett13b, degenaar11a, degenaar11b, fridriksson11}. 
The deep crustal heating model does not involve the optical/UV emission at all, and only applies to the thermal X-ray emission from the NS surface. However, in at least some quiescent LMXBs the X-ray flux is variable in a manner that can not be reconciled with model predictions.  Indeed, the thermal X-ray emission in several (though not all) systems seems to be highly variable during quiescence \citep{campana97, rutledge02, cackett10, cackett11}, which deep crustal heating alone cannot account for. Moreover, many quiescent NS LMXBs in addition to thermal emission, likely originating from the whole NS surface, also show a power law spectral component that cannot be explained by this model and whose origin is instead still unclear.

In the mid-1990s a different scenario for quiescent emission, called the advection dominated accretion flow (ADAF), was developed by \cite{narayan1} and \cite{narayan2}, and consisting in a set of slightly different models. For a review of ADAF models see for example \citet{narayanbook}. However, other models, always involving radiatively inefficient accretion flows, were also proposed, like in the case of the adiabatic inflow-outflow solution \citep[ADIOS][whereby a strong  outflow is driven]{blandford99}. More particularly, in the ADAF scenario, if the accretion rate drops below a critical value, the standard disc can become unstable and partially evaporate, resulting in a disc truncated further out from the compact object. The matter evaporated from the disc becomes an ADAF, assuming a spherical shape, and rotating much slower than the Keplerian velocity. The characteristic timescale of accretion is rapid, about $10\%$ of the free fall timescale, and is faster than the cooling timescale. This means that the accretion flow becomes radiatively inefficient. The energy released because of viscosity is consequently stored in the accretion flow itself (advection) as entropy (heat) and is not radiated away, unlike the case of a standard disc. Close to the compact object the accretion flow inefficiently cools through synchrotron emission from electrons and ions, producing UV radiation, while further out it is instead the bremsstrahlung which dominates, producing X-ray radiation. In the case of NS LMXBs only, Compton scattering between the seed photons arising from the NS surface and the hotter electrons in the flow, is the most efficient source of cooling. Finally, if the matter reaches the surface of the NS, all the energy in the flow is transferred to it and then radiated away, while it is lost inside of the event horizon, in the case of a BH.  This model predicts brighter X-ray emission from a NS, because its surface is heated by the flow, compared to a BH which instead advects matter and radiation. However, ADAF predictions fit the low quiescent X-ray luminosity of NS LMXB only with the extra action of an efficient propeller effect at the magnetospheric radius which prevents the most part of the accreting matter from reaching the NS surface \citep[see e.g.][]{asai98, menou99}. The development of an ADAF model able to make a clear prediction about NS LMXB O and UV emission is a real challenge. Indeed, the X-ray radiation emitted from the star surface strongly interacts with the accretion flow, which, consequently, cools down because of X-ray inverse Compton scattering. If the Compton cooling is efficient it could be difficult to observe the emission due to the ADAF, because it could become extremely faint \citep[e.g.][and references therein]{menou01}.

Another possibility, mainly proposed as an alternative to the ADAF plus propeller model, was developed by \cite{cam&stell}. They suggest that during quiescence the majority (if not all) of the mass transferred from the companion star is accumulated in the outer edge of the accretion disc, while just a tiny fraction of it (if any) is accreted on the compact object. For BH systems, the optical and UV emission is in excess of that  emitted by the companion star, is gravitationally released in the accretion disc. For NS system instead, the O and UV excess, together with the hard part of the X-ray photons (above 2 keV), are arising from the interaction of a radio pulsar relativistic wind with the matter transferred  from the companion, in a region where a shock front takes form. Since the pulsar emission is quenched from the accreting matter, no periodic signal is observed. The whole emission has a power law spectral shape, extending from the optical to the X-ray band.  However, in order to account for the emission of soft X-ray photons (below 2 keV), crustal cooling is needed or, alternatively, the magnetic pole of the NS surface must be hit and heated by relativistic particles moving along the magnetosphere.  Variation in the power law component of NS quiescent LMXBs, would support this model,  while variation in the thermal component would favour a scenario involving direct accretion.

Finally, another scenario was proposed to explain the UV emission, which in quiescence is supposed to mainly trace the accretion flow (the companion star is cold, and it is expected to have a minimal contribution at that energy). This involves irradiation and reprocessing of the X-ray emission from the accretion disc \citep[see e.g.][]{vanpar94}. However, reprocessing does not explain how X-rays are generated during quiescence. Summarizing, thus far, there is no clear picture of the quiescent accretion flows in LMXB, this is particularly true for that hosting a NS. Indeed, the fact that a NS possesses a surface and a magnetosphere makes it much more complicated than a BH. 

Despite much early observational and theoretical work on quiescent LMXBs, there are several crucial questions still remaining which only simultaneous multiwavelength campaigns can address.  For instance, is it accretion at low rates really happening onto NSs, and how exactly does it work? Where precisely does the UV emission arise from? Is the X-ray emission irradiating perhaps the inner edge of the accretion disc, leading to reprocessed UV emission, which would imply that X-ray variability is triggering UV variability?  Otherwise, are mass accretion rate fluctuations propagating inwards from the outer disc (UV), up to the NS surface (X-ray), implying that, on the contrary, UV variability is leading the X-ray variability?
In order to answer these questions and motivated by recent new evidence for an X-ray-UV correlation in Cen~X-4 \citep{cackett13}, we planned a new and unique study of the NS LMXB Cen X-4. Indeed, due to its proximity and the low level of extinction (see Sec. \ref{sec:constr}), Cen X-4 is the perfect target for a multi wavelength quiescent study.  For the first time ever, we intensively monitored the source on a daily basis, for more than two months, with simultaneous optical, UV, and X-ray observation performed by \Sw. 

In sect. \ref{sec:constr} we report on previous observations of Cen X-4 at which we will later refer in this work. In sect. \ref{sec:obs} we describe the \Sw\ observations and the data reduction. In sect. \ref{sec:results} we describe the data analysis and we show the results, that we later discuss in sect. \ref{sec:discuss}. In sect. \ref{sec:summary} we make a summary of the main work conclusions.

\section{What is known about Cen X-4} 
\label{sec:constr}

Cen X-4 was discovered in 1969 when it entered in an outburst state \citep{conner1969}. Ten year after it experienced a second outburst \citep{kaluzienski79} and it has been in quiescence from then on. Moreover, \citet{kuulk09} discussed the possibility of the detection of another X-ray burst from Cen X-4 detected by Apollo 15 in 1971. During the decay of the 1979 outburst, a X-ray type I burst was recorded, unambiguously showing that Cen X-4 hosts a NS \citep{matsuoka80}. Assuming that during the burst the NS (assuming to have a mass of $1.4M_{\odot}$) was emitting at the Eddington limit, an upper limit to its distance was derived and it is 1.2 kpc \citep{chevalier89}. The peak of the two outburst luminosity were $\sim10^{38}$ erg/s and $\sim0.2\times10^{38}$ erg/s.
Later, \cite{gonzalez05}, from the same burst luminosity, but comparing that with the average peak luminosities of photospheric radius expansion bursts observed in globular clusters, proposed a distance of $1.4\pm0.3$ kpc. Note that \citet{kuulk09} also derive an upper limit of 1.2 kpc from their reanalysis of a 1969 X-ray burst from Cen~X-4.  We assumed a distance of 1.2 kpc since this value matches all these results.

Cen X-4 is one of  the most studied LMXB at optical wavelength, since it is  the brightest (and nearest) quiescent NS LMXB at this energy.
Its mass ratio is well established and results $q=0.17\pm0.06$, while the mass of the NS and that of the companion are constrained to lie within $0.49<M_{NS}<2.49$ and $0.04<M_{c}<0.58$ \citep{torres02}. 
This implies, assuming that the secondary fills its Roche lobe, that the companion stellar
radius is $0.5<R_{c}/R_{\odot}<1.2$ \citep{gonzalez05}. The companion star is an evolved main sequence star of the K3-7 V type \citep{chevalier89, davanzo05} with an effective temperature T$_\mathit{eff}=4500\pm100$ K \citep{gonzalez05}. Photometric optical measure showed that its luminosity
is modulated at a period of 15.1 h, which has been interpreted as the orbital period. The modulation in the V bands has an amplitude of 0.2 magnitudes, which corresponds to a flux variation of about 17\%. 
The accretion disc, which has an inclination of $30-46^{\circ}$, is bright at optical wavelengths, 
producing $\sim25\%$ of the total light in the V band \cite{chevalier89}, while its contribution in the IR is lower, about 10\% \citep{shahbaz93}. By knowing the orbital period and the mass ratio, and assuming a NS of 1.4$M_{\odot}$, we derived some of the orbital parameters that we will refer to later: the orbital separation $a$ between the two stars of the binary is $\sim2.5\times10^{11}$ cm, the distance from the center of the companion and the inner Lagrangian point $b_{1}$ is $\sim1.7\times10^{11}$ cm, the outer disc radius is $\sim1.5\times10^{11}$ cm \citep{fkr92}.

During quiescence, the spectrum of Cen X-4 is well fitted by a two component model made by a power law, with spectral index $\alpha\sim1.4-1.8$, plus a NS atmosphere made of pure hydrogen of effective temperature $kT_\mathit{eff}=50-70$ eV, multiplied by a low level of interstellar absorption $N_{\rm H}=4.9\times10^{20}$ cm$^{-2}$ \citep{cackett13}. The X-ray light curve of Cen X-4 during quiescence is highly variable on a wide range of timescales, from hundreds of seconds to years, with count rate variations up to a factor of 3 in only four days. The 0.5--10 keV quiescent luminosity was found to be $\sim0.76-4.91\times10^{32}$ erg/s \citep{campana97, rutledge01, campana04, cackett10, cackett13}.

\section{Observations and data reduction}
\label{sec:obs}

The \Sw\ satellite has on board three telescopes covering respectively the optical/UV band, Ultraviolet/Optical Telescope, UVOT \citep{roming05, breeveld10}, the X-ray Telescope, XRT, and the gamma-ray band, Burst Alert Telescope, BAT \citep{gehrels04}. Because of the soft nature of the X-ray emission from Cen X-4, we only analyzed data from the XRT, covering the 0.3--10 keV energy range, while the source is undetected in the BAT range. Concerning the optical/UV analysis we explored data in 2 optical filters, V (5468\AA) and B (4392\AA) and 4 UV filters U (3465\AA), UVW1 (2600\AA), UVM2 (2246\AA) and UVW2 (1928\AA).

\Sw\ observed Cen X-4 for a total of 62 times\footnote{The observation id number goes from 00035324001 to 00035324065 with the exception of 00035324037/49/64} (60 times in PC mode), 58 of this pointing are performed on a daily basis between 5 June 2012 and 8 August 2012. The 60 observations in PC mode that we used were performed in the following way: one 5 ks pointing was followed by 4 pointings of 1 ks each and so on. The UVOT configuration was set as follows: for the observations lasting 5 ks the UVW2, UVM2, UVW1, U, B, V filters are sequentially used, while for the observation lasting 1 ks, only the UVW1 and V filters are used. Each observation made by \Sw\ is always composed of multiple snapshots, both in the case of XRT and UVOT, the total number of which is variable among different observations. For each single XRT snapshot, multiple UVOT (sub)snapshots are made, one for each of the six available bands (UVW2, UVM2, UVW1, U, B, V) in case of 5 ks observation, while one for the UVW1 and one for the V band, in case of 1 ks observation.   We first analyzed the average data of each single observation, summing together all the XRT snapshots and all the (sub)snapshots of the same UVOT filter. Consequently, each point in the X-ray light-curve has an exposure time of 1--5 ks, while each pointing in the O-UV light-curve lasts from few hundred seconds up to about 2 ks. Secondly, we separately analyzed the data of all single UVOT (sub)snapshot, then extracting only the part of the XRT snapshot exactly simultaneous to the UVOT one. This procedure results in single exposures ranging from about few tens of seconds up to about 2 ks.
 
\subsection{XRT data reduction}

In order to perform an accurate data analysis of the source variability with \Sw\ data, it is mandatory to remove the contamination from bad pixels and bad columns which is affecting the XRT CCD after it was hit by a micrometeorite in 2005. We followed the detailed online thread procedure available at 
http://www.swift.ac.uk/analysis/xrt/xrtpipeline.php, that we summarize below.

Using the $HEAsoft$ verision 6.12\footnote{http://heasarc.gsfc.nasa.gov/lheasoft/} we first produced the cleaned events-file of each of the 60 single pointings in PC mode with the \textit{xrtpipeline} command by also creating the pointing exposure map, manually providing the best source position. Using the cleaned event file, we produced the source image and then extracted the source spectrum from a circular region of 20 pixel size, centered at the source position. The background spectrum was extracted in a clean region of the sky near the source, from a region of 80 pixel size. Due to the low source X-ray flux ($\sim0.006-0.021$ c/s), no pile-up correction is needed in the extraction of the source spectrum.  The ancillary response file (ARF) was then created, referring to the exposure map previously generated, by using the \textit{xrtmkarf} tool, with the PSF correction active which, among other things, ensures a correction for hot columns and bad pixels. 

We also produced the cleaned event-file for each XRT snapshot of each observation and, on a shorter timescale, also for all the multiple sub-intervals, lasting a few hundred seconds on average, corresponding to the time interval of effective UVOT snapshot exposure. In order to extract the source spectrum for each time interval of interest, we followed the procedure described above, producing a single exposure map for each of the time intervals. Finally, we calculated the source background subtracted count rate from the spectral files taking into account the ratio between the area of the source region and background region. 

In order to create spectra with the maximum possible signal to noise ratio (S/N) we also summed together spectra from different XRT pointing, with comparable count rate, using the following procedure. We first sum all the individual event files of interest by using \textit{XSelect}, then we summed, one by one, the individual exposure maps using the command \textit{sum ima} in the \textit{Ximage} package. We finally extracted the summed spectrum with \textit{XSelect} and generated the summed ARF file using the summed exposure maps. 
Before fitting, the summed spectra were rebinned in order to have, at least, 20 counts per bin, and fits are performed using the $\chi^2$ statistic. All spectral fits are made with XSPEC version 12.7.1 \citep{arnaud96}, and the latest calibration files available in October 2012.

\subsection{UVOT data reduction} 

We applied the following procedure consisting of five consecutive steps (each one using a specific tool) to each observation and each band filter contained in it.  The five steps are: 1) \textit{uvotbadpix}, 2) \textit{uvotexpmap}, 3) \textit{uvotimsum}, 4) \textit{uvotdetect}, 5) \textit{uvotsource}.

In more detail, starting from the raw image file, we first produced the bad pixel map with the \textit{uvotbadpix} tool. As a second step, from the level II image fits file, we created the exposure maps, by using the bad pixel map previously generated as reference for bad pixels positions, and the \textit{uat.fits} file as attitude file (more accurate than the \textit{sat.fits} file). We summed all the multiple image extensions contained in the level II image fits file, producing the total image, by using the \textit{uvotimsum} tool. We did the same for the exposure map, by using the \textit{uvotimsum} tool we built the total exposure map. We run the \textit{uvotdetect} tool to detect all the source in the summed image above a $3\sigma$ threshold. We identified the closest source to the Cen X-4 coordinates in the output \textit{uvotdetect} list file, also checking that the source was always detected. Finally, we extracted the background subtracted source count rate by using the \textit{uvotsource} tool, after defining the source and the background regions with a circle of 5 and 10 arcs size respectively. The source region is centered at the best position known for Cen X-4, while the background region is taken from a close by region, not contaminated by other sources.

We extracted the source count rate for each single snapshot, of each band, for all the pointings, following the same procedure described above, except for the fact that the exposure maps and the image files, were not summed together (step 3 of the total reduction procedure was skipped as in this case we are only looking at individual exposures). 

\section{Analysis and results}
\label{sec:results}

\subsection{Multiwavelength light curve}

\begin{figure*}
\begin{center}
\includegraphics[angle=-90,width=5.0in]{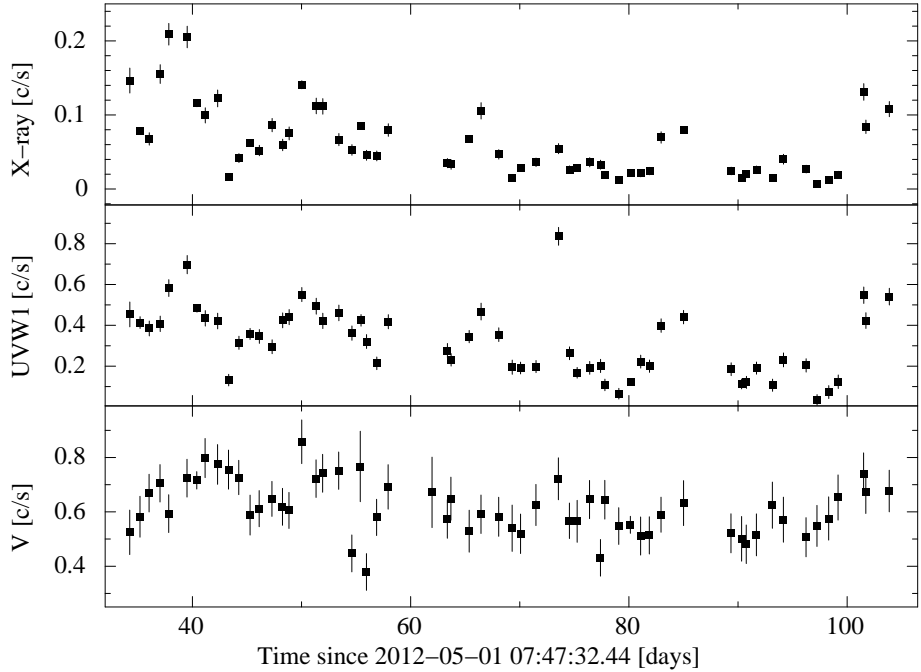} \\ 
\caption{Light curves for Cen X-4 in three different energy ranges.  0.3--10 keV, X-ray (top),  
UVW1, 2600\AA\ (center) and V, 5468\AA\ (bottom). For plotting purposes we selected 
the reference time as 2012-05-01 07:47:32.44 UTC, which corresponds to the second \Sw\ pointing. 
The only two pointings which are not plotted here correspond to t$=0$~d and t$=-2066.8$~d.}
\label{fig:3b_lc.ps}
\end{center}
\end{figure*}

We first analyzed the background-subtracted light curves in all the energy bands. The source is highly variable in all bands. Particularly interesting is the comparison between the X-ray and the UVW1 and V bands, the three energy intervals for which the most pointings (60) are accumulated. Indeed, due to the selected observing strategy, \Sw\ was always observing in those bands. We show the light curves in Fig. \ref{fig:3b_lc.ps}. For plotting purposes we do not show the two first pointings (t=-2066.8 d, and t=0, where t=0 is 2012-05-01 07:47:32.44 UTC) that do not belong to the main observational campaign.  
The X-ray and UVW1 light curve seem to follow roughly the same pattern, showing both long term (months) and short term (days) variability. They both show randomly distributed peaks in the count rate at the same time. The V light curve is  much less variable, however, the uncertainties here are larger. We note that the recorded peaks are independent from the orbital period which is only $\sim15.1$~h \citep{chevalier89}. \citet{zurita02} also measured flare-like optical variability from Cen X-4 showing that it is independent from the orbital period.

We calculated the fractional root mean square variability (F$_{\rm var}$) for the light curves in Fig. \ref{fig:3b_lc.ps} \citep[see][for more detail about F$_{\rm var}$]{vaughan05}.  
All uncertainties are hereafter at the $1\sigma$ confidence level.
F$_{\rm var}$ is $73.0\pm1.5\%$ for the X-ray band,  $50.0\pm1.4\%$ for the UVW1 band, and $10.0\pm1.6\%$ for the V band, implying that the X-ray emission is the most variable. The most significant changes in the X-ray count rate are detected between 39.5 d and 43.3 d and between 97.2 d and 101.5 d,  where the count rate respectively decreases by a factor of $\sim13$ and increases by a factor of $\sim22$ in four days only (from $0.21\pm0.01$ c/s to $0.016\pm0.04$ c/s, and from $0.006\pm0.003$ c/s to $0.13\pm0.01$ c/s respectively).  In the same time window, the UVW1 emission is changing less, and decreases by a factor of $\sim5$ (from $0.70\pm0.04$ c/s to $0.13\pm0.03$ c/s) and increases by a factor of $\sim14$ (from $0.04\pm0.03$ c/s to $0.55\pm0.04$ c/s) respectively.

\subsection{Structure Function}

The first order structure function is commonly used to establish the timescale and the intrinsic variability in the light curve of AGN \citep[see e.g.][and reference therein]{do09}. The structure function is related to the autocorrelation function and the power spectrum and is an alternative way of determining the power at a given timescale.  In our case we have a light curve that is not evenly sampled, thus determining the power spectrum becomes complicated.  The structure function provides us with a straight-forward method in the time domain.

\begin{figure}
\centering
\includegraphics[angle=-90,width=8.5cm]{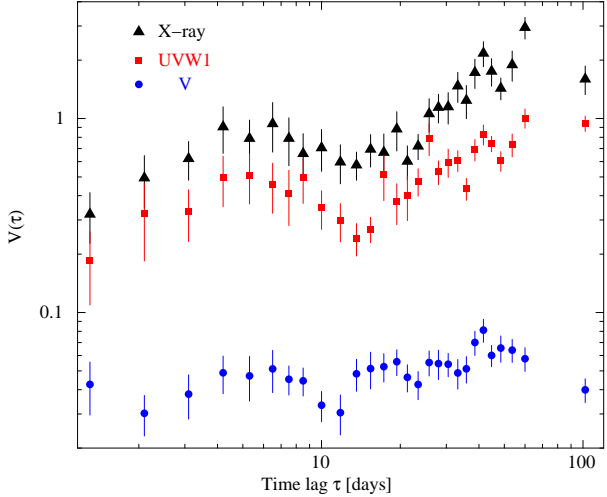}
\caption{First order structure function $V(\tau)$ as a function of the time lag for the three light curves (X-ray, UVW1, and V) in Fig. \ref{fig:3b_lc.ps}. The light curves have been normalized to their average value before computing the structure function. This allows for easy comparison of the variability amplitude in each band.}
\label{fig:structure}
\end{figure}

Here, we used (for the very first time) the structure function to define the quiescent variability of Cen X-4. We compute the structure function for the three light curves (X-ray, UVW1, and V) in Fig. \ref{fig:3b_lc.ps}. The structure functions are computed after having normalized the light curves to their average values (this allows for a direct comparison of the intensity of the variability among the three light curves). For this structure function analysis we also included the point at $t=0$ in the lightcurve (not shown in Fig.~\ref{fig:3b_lc.ps}, so the light curve covers a total of $\sim104$ days. The first-order structure function is here defined as:
\begin{equation}
V(\tau) = \langle[s(t+\tau)-s(t)]^{2}\rangle
\end{equation}
where the bracket denotes an average quantity, $s(t)$ is a set of measures (the count rate in each band) performed at the time $t$. Since our data is unevenly sampled, we measured $[s(t+\tau)-s(t)]^{2}$ for all the existing couples of time lags.  Then we binned the time lags, with a variable bin size in order to  have a comparable number of points in each bin with a minimum number of points equal to 40. The center of each bin is the time lag for that bin, while the average of  $V(\tau)$ is the value of the structure function for each bin at that time lag. The error associated to each bin is the standard deviation of the structure function $V(\tau)$ divided by the square root of the number of points in the bin ($\sigma_{bin}/\sqrt{N_{bin}}$). The three structure functions are showed in Fig. \ref{fig:structure}. The shape of an ideal structure function is expected to follow a shape consisting of two plateaus, the first one at the timescale of the noise and the second at a timescale longer than the inspected physical process, connected to each other by a power-law which provides a measure of the timescale of the variability \citep[see, e.g.][for a clear description and example]{hughes92}.  The slope of the structure function is related to the slope of the power spectrum.  For instance, a power spectrum with $P(f) \propto f^{-2}$ will show a first-order structure function where $V(\tau) \propto \tau^{1}$ \citep[e.g.,][]{hughes92}.  We find the power law index of the structure function of Cen X-4 to be $1.0\pm0.1$ and $0.8\pm0.1$ and $0.3\pm0.1$ in the case of the X-ray, UVW1, and V respectively, for $11<\tau<60$. The fit was performed in the time lag range of 11--60 days (where the power law extends) and gives $\chi^{2}_{\nu}=1.36$, $\chi^{2}_{\nu}=1.70$, and $\chi^{2}_{\nu}=0.86$ (16 d.o.f.) respectively. The slope of the X-ray and UVW1 band is consistent with a power spectrum with index $-2$ (red noise also known as random walk noise) as typically seen in accreting systems, while the slope of the V band is different and, consequently, it may be associated with another underlying process.
The structure function also shows that the amplitude of the variability is greater in the X-ray band than in others, and the UVW1 amplitude is greater than in the V band (matching the results of the $F_{var}$ analysis). The timescale of the variability goes from 10 days up to at least 60 days (there is only one point in the light curve a longer timescale). However, an excess is present at about 4-5 days. This is likely linked to the rise and decay time of the peaks in the light curve. So, variability is also detected at this timescale.

\subsection{Long timescale correlation}

\begin{figure*}
\begin{center}
\includegraphics[angle=-90,width=7.0in]{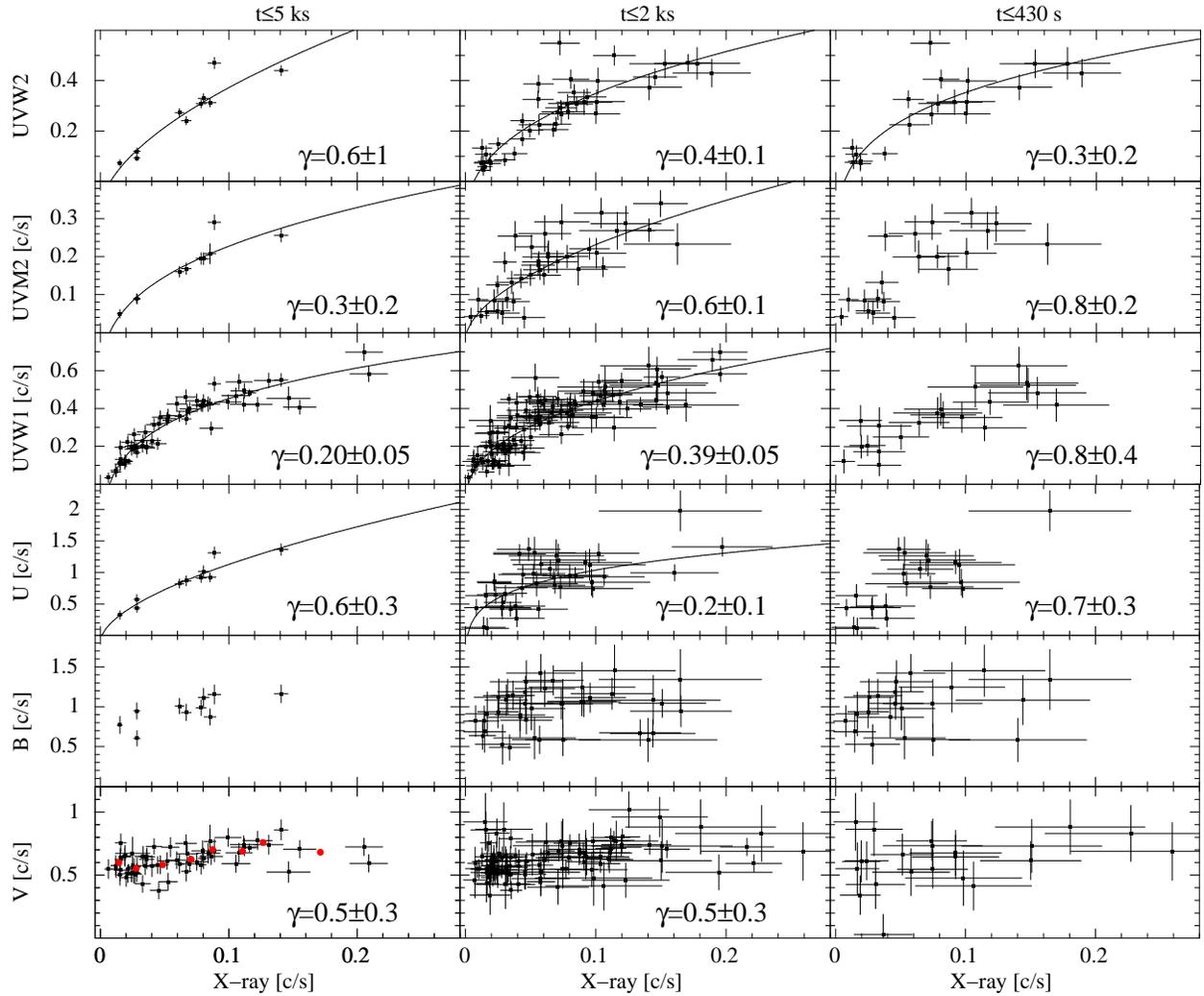} \\ 
\caption{Study of the X-ray vs UV-O correlation as a function of the energy interval and the timescale. Energy increases from bottom to top, while the timescale decreases from left to right. The solid line represents the best fitting model, a constant plus a power law, which is plotted only when it is unambiguously determined. The value of the slope $\gamma$ for the constant plus power law model is also reported in the lower right part of each panel. A correlation is always found, exception are the B band (all timescale), where, however, no high X-ray count rate coverage is available, and the V band (only in the case of $t\leq110$ s), likely because of the limited number of points. \textit{Left:} X-ray vs UV-O average count rates on the XRT pointing timescale (t$\leq5$ ks).  For plotting purposes, the V band count rate is also averaged (red circles) in 8 X-ray count rate sub intervals (0--0.002, 0.002--0.004, 0.004--0.006, 0.006--0.008, 0.008--0.010, 0.010--0.012, 0.012--0.022 c/s). \textit{Center:} X-ray vs UV-O count rate on the UVOT snapshot timescale (t$\leq2$ ks). \textit{Right:} X-ray vs UV-O count rate on the UVOT shorter snapshot timescale t$\leq430$ s (here the exact timescale depends on the energy interval. See text for more details).}
\label{fig:corr_5ks}
\end{center}
\end{figure*}

We studied the correlation between the X-ray count rate and that of all UV and optical bands on a timescale corresponding to the XRT exposure, t$=1-5$ ks (t$\leq5$ ks hereafter). This means to compare the average count rate of each XRT pointing with that of the simultaneous O and UV bands summed snapshots (see Fig. \ref{fig:corr_5ks}). In order to prove if a correlation is present, we first fit the data with a constant, then with a constant plus a linear ($y=y_{0}+ax$) and with a constant plus a power law ($y=y_{0}+ax^{\gamma}$). We verified the statistical significance of the inclusion of the linear and the power law component with respect to a constant with an F-test.  If they are significant at more than the $3\sigma$ confidence level, we consider that a correlation exists.  
Then, we verified if the shape of the correlation is a linear or a power law function. 
Since we are comparing two different models, with a different number of $dof$, 
we can not use an F-test. Consequently, we applied the following procedure:
we compared the constant plus linear fit with that of a constant plus power law,  by estimating the $\chi^{2}$ distance, here defined as $x$, from its expectation value in unit of $\sigma_{\rm \chi^2}$.
Since $\sigma^{2}_{\rm \chi^{2}}= 2~dof\Rightarrow\sigma_{\rm \chi^{2}}=\rm \sqrt{2{\it dof}}$, and so
  $\frac{(\chi^{2}-dof)}{\sigma_{\rm \chi^{2}}}= x[\sigma_{\rm \chi^2}]$,
where $dof$ are the degrees of freedom.  We consider as the most likely shape of the correlation that having the lower value of $x$. All fit results are reported in Table~\ref{tab:corr5000s}.  Note that for the UVW1 band, we excluded from the correlation study the anomalously high count rate ($\sim0.85$ c/s) point at t=75.5 d (obsid 00035324038), since it evidently lies significantly outside of the correlation. Moreover, for that observation no coverage is provided in the other UV bands, and so it is not possible to cross check if such high count rate is also present in there. We note that optical flares not associated with the X-ray source, but likely linked to some mechanism in the companion star or in the disc, have been already observed from Cen X-4 \citep{zurita02}.

We always found a correlation between X-ray and all the UV bands (UVW2, UVM2, UVW1, U), see Fig. \ref{fig:corr_5ks}. Although some scatter is always present, the fit with a constant plus a power law is always preferred in terms of $x$. 
The significance for the inclusion of a power law in the fit ranges from 3.6$\sigma$ up to more than 8$\sigma$ in the case of the UVW1, where more data points are available. The value of the power law index $\gamma$ ranges from $0.20\pm0.05$ (UVW1) to $0.6\pm0.3$ (U). No correlation is detected in the case of the optical B band (the significance for the inclusion of a power law is only $1.6\sigma$, see also Tab. \ref{tab:corr5000s}). However, here, only 9 points are available, moreover, none of them is covering the high count rate region above 0.17 c/s, where a power law and a linear function clearly reveal their difference.  A correlation with the V band is also found.  Either a power law and a linear function, both significant at more than $4\sigma$, provide comparable fit in term of $x$.

Note that \cite{cackett13} found a linear correlation between the X-ray and the UV (UVW1 \XMM\ band) in Cen X-4, while we find here a power law to be the best-fitting shape of the correlation. This can easily been explained by the fact that we have a significantly higher number of data points and, consequently, we are covering a wider dynamic range in X-ray count rate, especially at high X-ray count rate (e.g. above 0.17 c/s), where a linear function and a power law really show their difference in shape.

\subsection{Short, and very short timescale correlation}

In order to further explore the presence and the nature of a correlation between X-ray and all UV-O bands emission, we compared the count rates on shorter timescale. We selected that of the single UVOT snapshot inside of a \Sw\ observation, where $t$ is going from $\sim10$ s up to $\sim2$ ks, so the explored timescale here is t$\leq2$ ks. In this case a higher number of data points is available, ranging from $\sim40$ up to $\sim110$, depending on the band (see Fig. \ref{fig:corr_5ks}). Finally, we investigate the correlation on the shortest timescale available by selecting the shortest UVOT snapshot only, with the condition to maintain a statistically relevant number of data points (n$\geq20$). The correlation on such short timescales was never studied before.

Using the above conditions, we explored, for the different bands, the following timescale: $t\leq$430 s for UVW2, $t\leq$320 for UVM2, $t\leq$150 s for UVW1, $t\leq$110 s for U, $t\leq$110 s for B, and $t\leq$110 s for V. Consequently, we define the very short timescale as $t\leq$430 s. Although, once again, some scatter is always present, as in the case of t$\leq5$ ks, a correlation between the X-ray and all UV bands is always found, for both t$\leq2$ ks and t$\leq430$ s. This is the very first time that a correlation is observed at that timescale.
In the case of t$\leq2$ ks (see Tab. \ref{tab:corr1500s}), the most likely shape of the correlation is always a power law, as $x$ is always lower than in the case of the fit with a linear function. The significance of the inclusion of the power law function with respect to a constant is always greater than $4\sigma$. We found $\gamma$ ranging from $0.2\pm0.1$ in the case of U band up to $0.6\pm0.1$ for the UVM2 band. No correlation is detected with the B band, while a correlation is found with the V band, where both a linear and a power law shape are equally acceptable from a statistical point of view. In the case of $t\leq430$ s (see Tab. \ref{tab:corr110s}), for the UVW2 band, the most likely shape of the correlation is a power law ($\gamma=0.3\pm0.2)$, while it could be either a power law or a linear for the UVM2, UVW1 and U bands. However, we note that in the latter bands, due to the limited number of data points, the explored X-ray count rate is always lower than 0.17 c/s. No correlation is detected either with the B and the V band (likely because of the limited number of data points and because the high X-ray count rate is not covered). 

Summarizing, we studied the presence of a correlation between the X-ray count rate and that of UV and optical bands (6 in total) for three timescales ($t\leq5$ ks, $t\leq2$ ks, $t\leq430$ s). We found a statistically significant correlation in 14 cases over a total of 18 (the exception are the B band, for all timescale, and the V band, for the very short timescale only). In 9 cases over a total of 14 where a correlation is significant, its most likely shape is a power law. A linear shape is instead disfavoured (see Fig. \ref{fig:corr_5ks} and see the value of $x$ in Tab. \ref{tab:corr5000s},\ref{tab:corr1500s}, and \ref{tab:corr110s}).
Finally, we cross checked the results of the above correlation study, for all timescale, with a Spearman's correlation test. We reported in Tab. \ref{tab:corr5000s}, \ref{tab:corr1500s}, and \ref{tab:corr110s} the value of the Spearman's coefficient $\rho$ and the null hypothesis probability ($P$), which is the probability that any random sample of uncorrelated experimental data points, of an uncorrelated parent population, would yield a Spearman's coefficient value equal to $\rho$. The results of the Spearman's test fully confirm our first results. There is always a strong correlation between the X-ray and UV emission in all bands for all timescale, a less intense but still statistically significant correlation is also found with the optical V band emission (for $t\leq5$ ks and $t\leq2$ ks). No statistically significant correlation is detected with the V band for $t\leq110$ s and with the B band (all timescale), very likely because of the limited number of data points (indeed the probability P is strongly dependent on the number of data point).   

\begin{table*}
\caption{Study of the correlation between X-ray and UV and O bands on the timescale of an XRT observation, t$\leq5$ ks. 
The results of the fit with a linear (co+li) and a power law (co+pow) are reported only when these component are statistically significant.
$Co$, $Li$, $\gamma$ and $a$ are the value of the constant, the slope of the linear, and the index of the power law with its multiplicative coefficient $a$ respectively, while $\rho$ is the Spearman's coefficient and P the null hypothesis probability (no correlation). Uncertainties are $1\sigma$ confidence level.}
\begin{center}
\begin{tabular}{cccccccccccc}
\hline 
Band & Model    &  $Co$           &      $Li$     & $\gamma$      &    $a$       & $\chi^2_{\nu}$/dof  & sign.       & $x$        &    $\rho$ & $P$   \\
     &          &                 &               &               &              &               & $\sigma$    & $\sigma_{\rm \chi^{2}}$      \\ 
\hline       
UVW2 & co       & $0.22\pm0.05$   &               &               &              & 69.7/9        &             & 145.8    &     &  \\
 -   & co$+$li  & $0.02\pm0.01$   & $3.5\pm0.1$   &               &              &  7.9/8        & 4.0         &  13.9    &  0.964   & $7.3\times10^{-6}$ \\
 -   & co$+$pow & $-0.11\pm0.06$  &               & $0.6\pm0.1$   & $1.8\pm0.3$  &  7.3/7        & 3.6         &  11.7    &     &        \\
                                                                                                                    
UVM2 & co       & $0.14\pm0.05$   &               &               &              & 25.9/9        &             & 52.9     &     &   \\
 -   & co$+$li  & $0.04\pm0.01$   & $1.9\pm0.1$   &               &              &  3.6/8        & 3.8         &  5.2     & 0.964    &   $7.3\times10^{-6}$      \\
 -   & co$+$pow & $-0.2\pm0.1$    &               & $0.3\pm0.2$   & $0.8\pm0.1$  &  2.6/7        & 3.7         &  2.9     &     &         \\
                                                                                                                    
UVW1 & co       & $0.284\pm0.004$ &               &               &              & 25.1/58       &             & 129.6    &     &   \\
 -   & co$+$li  & $0.109\pm0.006$ & $3.3\pm0.1$   &               &              &  5.0/57       & $>8$        &  21.1    & 0.934   &  $3.6\times10^{-27}$       \\
 -   & co$+$pow & $-0.7\pm0.3$    &               & $0.20\pm0.05$ & $1.8\pm0.3$  &  2.3/56       & $>8$        &   7.0    &     &         \\
                                                                                                                    
U    & co       & $0.80\pm0.02$   &               &               &              & 18.2/9        &             & 36.5     &     &    \\
 -   & co$+$li  & $0.25\pm0.04$   & $9\pm1$       &               &              &  2.0/8        & 4.1         &  1.9     & 0.976    &   $1.5\times10^{-6}$      \\
 -   & co$+$pow & $-0.1\pm0.3$    &               & $0.6\pm0.3$   & $5\pm2$      &  1.8/7        & 3.6         &  1.6     &     &         \\
                                                                                                                    
B    & co       & $0.94\pm0.3$    &               &               &              &  2.8/9        &             &  3.8     &     &        \\
 -   & co$+$li  &                 &               &               &              &  1.4/8        & 2.2         &          & 0.697    &  $0.025$        \\
 -   & co$+$pow &                 &               &               &              &  1.6/7        & 1.6         &          &     &         \\
                                                                                                                    
V    & co       & $0.61\pm0.01$   &               &               &              &  2.0/59       &             & 5.3      &     &    \\
 -   & co$+$li  & $0.55\pm0.02$   & $1.1\pm0.2$   &               &              &  1.4/58       & 4.4         & 2.2      & 0.554    &   $4.3\times10^{-6}$       \\
 -   & co$+$pow & $0.5\pm0.2$     &               & $0.5\pm0.3$   & $0.6\pm0.3$  &  1.4/57       & 4.2         & 2.1      &     &          \\
\hline 
\\
\end{tabular}  
\label{tab:corr5000s}                                                                                                         
\end{center}
\end{table*}

\begin{table*}
\caption{Study of the correlation between X-ray and UV-O bands for the timescale of an UVOT snapshot, t$\leq2$ ks. 
See Tab. \ref{tab:corr5000s} for the definitions of the parameters here reported.}
\begin{center}
\begin{tabular}{ccccccccccc}
 \hline 
Band & Model    &  $Co$           &      $Li$     & $\gamma$      &    $a$        & $\chi^2_{\nu}$/dof  & sign.       & x &    $\rho$ & $P$ \\
     &          &                 &               &               &               &               & $\sigma$    & $\sigma_{\rm \chi^{2}}$ \\ 
\hline              
UVW2 & co       & $0.217\pm0.005$ &               &               &               & 18.1/39       &             & 75.7  &     &    \\
 -   & co$+$li  & $0.056\pm0.008$ & $2.8\pm0.1$   &               &               &  4.1/38       & $>8$        & 13.6  &  0.876   &  $1.4\times10^{-13}$  \\
 -   & co$+$pow & $-0.18\pm0.08$  &               &$0.4\pm0.1$    & $1.3\pm0.1$   &  3.3/37       & $>8$        &  9.8  &     &    \\
                                                                                                                                  
UVM2 & co       & $0.144\pm0.005$ &               &               &               &  7.7/36       &             & 28.3  &     &    \\
 -   & co$+$li  & $0.051\pm0.08$  & $1.8\pm0.1$   &               &               &  2.5/35       & $>8$        &  6.4  &  0.817   &  $6.9\times10^{-10}$  \\
 -   & co$+$pow & $-0.02\pm0.05$  &               &$0.6\pm0.1$    & $0.9\pm0.2$   &  2.4/34       & $>8$        &  5.7  &     &    \\
                                                                                                                                  
UVW1 & co       & $0.283\pm0.004$ &               &               &               & 14.6/108      &             & 99.6  &     &    \\
 -   & co$+$li  & $0.125\pm0.006$ & $3.0\pm0.1$   &               &               &  4.3/107      & $>8$        & 24.2  &  0.849   & $2.0\times10^{-31}$   \\
 -   & co$+$pow & $-0.12\pm0.05$  &               & $0.39\pm0.05$ &$1.38\pm0.06$  &  3.4/106      & $>8$        & 17.5  &     &    \\
                                                                                                                                  
U    & co       & $0.79\pm0.02$   &               &               &               &  4.9/39       &             & 17.3  &     &   \\
 -   & co$+$li  & $0.50\pm0.04$   & $5.0\pm0.6$   &               &               &  3.1/38       & 4.1         &  9.3  &  0.648   &  $6.5\times10^{-6}$   \\
 -   & co$+$pow & $-0.8\pm0.1$    &               & $0.2\pm0.1$   & $2.9\pm0.2$   &  3.0/37       & 4.0         &  8.5  &     &    \\
                                                                                                                                  
B    & co       & $0.94\pm0.03$   &               &               &               &  1.5/39       &             &  2.1  &     &   \\
 -   & co$+$li  &                 &               &               &               &  1.5/38       & $<0.5$      &  2.2  &  0.262   &  $0.103$   \\
 -   & co$+$pow &                 &               &               &               &  1.5/37       & $<0.5$      &  2.2  &     &    \\
                                                                                                                                  
V    & co       & $0.62\pm0.01$   &               &               &               &  1.6/102      &             &  4.5  &     &   \\
 -   & co$+$li  & $0.55\pm0.02$   & $1.0\pm0.2$   &               &               &  1.3/101      & 4.6         &  2.3  &  0.391   & $4.8\times10^{-5}$    \\
 -   & co$+$pow & $0.5\pm0.2$     &               & $0.5\pm0.3$   & $0.6\pm0.3$   &  1.3/100      & 4.3         &  2.3  &     &    \\
\hline 
\\
\end{tabular}  
\label{tab:corr1500s}                                                                                                         
\end{center}
\end{table*}
  
\begin{table*}
\caption{Study of the correlation between X-ray and UV and O bands for the shortest UV snapshot only. 
Here the explored timescale depends on the energy interval and it is consequently reported for each band. 
See Tab. \ref{tab:corr5000s} for the definitions of the parameters here reported.}
\begin{center}
\begin{tabular}{cccccccccccc}
 \hline 
timescale   &Band & Model     &  $Co$           &      $Li$     & $\gamma$      &    $a$             & $\chi^2_{\nu}$/dof  & sign.         & x &    $\rho$ & $P$ \\
    s       &     &           &                 &               &               &                    &               & $\sigma$      & $\sigma_{\rm \chi^{2}}$ \\ 
\hline
 430        &UVW2 & co        &  $0.23\pm0.01$  &               &               &                    &  14.6/19      &               & 42.0  &     &    \\   
            & -   & co$+$li   &  $0.07\pm0.02$  & $2.6\pm0.2$   &               &                    &   4.7/18      & 4.4           & 11.1  &  0.789   &  $3.5\times10^{-5}$  \\  
            & -   & co$+$pow  &  $-0.4\pm0.3$   &               & $0.3\pm0.2$   & $1.3\pm^{1}_{0.1}$ &   3.8/17      & 3.7           &  8.3  &     &    \\  
                                                                                                                                                       
 320        &UVM2 & co        &  $0.13\pm0.1$   &               &               &                    &   7.7/20      &               & 20.0  &     &    \\
            & -   & co$+$li   &  $0.05\pm0.02$  & $2.0\pm0.2$   &               &                    &   2.8/18      & 3.7           &  5.4  & 0.759    &  0.0001  \\  
            & -   & co$+$pow  &  $0.02\pm0.03$  &               & $0.8\pm0.2$   & $1.4\pm0.7$        &   2.6/17      & 3.6           &  5.6  &     &    \\  
                                                                                                                                                       
 150        &UVW1 & co        &  $0.31\pm0.02$  &               &               &                    &   3.7/20      &               &  8.5  &     &    \\  
            & -   & co$+$li   &  $0.14\pm0.03$  & $2.4\pm0.3$   &               &                    &   1.1/19      & 4.8           &  0.2  & 0.813    &  $7.5\times10^{-6}$  \\  
            & -   & co$+$pow  &  $0.10\pm0.07$  &               & $0.8\pm0.4$   & $2\pm1$            &   1.1/18      & 4.3           &  0.3  &     &    \\  
                                                                                                                                                       
 110        &U    & co        &  $0.71\pm0.04$  &               &               &                    &   5.5/20      &               & 14.1  &     &    \\
            & -   & co$+$li   &  $0.26\pm0.07$  & $9\pm1$       &               &                    &   2.8/19      & 3.4           &  5.7  & 0.668    &  0.0009  \\  
            & -   & co$+$pow  &  $0.1\pm0.4$    &               & $0.7\pm0.3$   & $5\pm3$            &   2.9/18      & 3.0           &  5.8  &     &    \\  
                                                                                                                                                       
 110        &B    & co        &  $0.97\pm0.05$  &               &               &                    &   1.4/20      &               &  0.8  &     &   \\
            & -   & co$+$li   &                 &               &               &                    &    1.3/19     & $<0.5$        &  0.7  & 0.347    & 0.1236   \\ 
            & -   & co$+$pow  &                 &               &               &                    &    1.3/18     & $<0.5$        &  0.9  &     &    \\ 
                                                                                                                                                       
 110        &V    & co        &  $0.59\pm0.04$  &               &               &                    &    1.3/21     &               &  0.9  &     &    \\
            & -   & co$+$li   &                 &               &               &                    &    1.2/20     & 1.2           &  0.6  & 0.238    & 0.2864   \\   
            & -   & co$+$pow  &                 &               &               &                    &    1.2/19     & 0.5           &  0.5  &     &    \\  
\hline  
\\
\end{tabular}  
\label{tab:corr110s}                                                                                                         
\end{center}
\end{table*}

\subsection{X-ray Spectral analysis}
\label{subs:spec}

As a first step we verified that the average spectral shape is not clearly changing among different pointings with comparable count rate. We visually inspected the spectral shape of each pointing, then we fitted the spectra with a simple model consisting of a power law multiplied by the Galactic photoelectric absorption \textit{phabs}. We imposed \textit{phabs} to be constant between observations  and we left the power law free to vary. The power law photon index was constant within statistical uncertainty, while only the normalization (flux) was found to change. We concluded that the spectral shape is roughly constant among different pointing with comparable count rate. Then, in order to inspect possible spectral change as a function of the flux (count rate), we selected three count rate range: low $<0.07$ c/s, medium 0.07--0.11 c/s, and high $>0.11$ c/s (see Fig. \ref{fig:3b_lc.ps}),  and we produced three summed spectra. This choice allowed us to maintain a comparable number of count among the three spectra. The low state spectrum has $\sim2100$ counts for 67 ks of exposure, the medium state spectrum $\sim2300$ counts for 30 ks, and the high state $\sim1400$ counts for 11.5 ks.

We selected a model made by the sum of a NS atmosphere composed of pure hydrogen \textit{nstamos} \citep{heinke}, where the magnetic field is assumed negligible (less than 10$^9$ G), plus a power law, both multiplied by \textit{phabs}, which accounts for photoelectric absorption due to the interstellar medium.  
We note that the use of the \textit{nstamos} model is not formally correct if the source is still accreting also at low Eddington luminosity rates. This is a scenario that we want to test in the following for Cen X-4. \cite{zampieri95} developed a model called $zamps$, explicitly to reproduce the thermal emission from a non magnetized NS, accreting at low rates ($10^{-7}\lesssim L/L_{Edd} \lesssim 10^{-3}$). However, \cite{zampieri95, soria11} showed that the fit performed with the $zamps$ model and that with the $nsa$ model, describing a NS with hydrogen atmosphere, are virtually identical for the level of S/N of our data. Consequently to enable comparison with previous work, we decided to use the \textit{nstamos} model.

Previous analysis showed that both the thermal component and the power law component must vary. In the case of the thermal component either the radius or the temperature can change, providing equally acceptable fits. Here, we decided to first fix the NS radius at 10 km, initially imposing that all the surface is emitting (normalization fixed to one). We fixed the source distance to 1.2 kpc assuming a NS mass of 1.4 M$\odot$. We fitted the three spectra together.

We first verified that the spectral changes among the three different count rate spectra, high (h), medium (m), low (l), can not be due to a change in the $N_{\rm H}$ alone. Consequently, we imposed the temperature of the NS surface, the power law slope $\Gamma$ and its normalization to be the same between different spectra, but free to vary. We get $\chi^2_{\nu}=2.1$ for 193 dof. We conclude that a change in $N_{\rm H}$ alone can not account for the observed spectral change. Then, we verified that the spectral changes can not be due to the variation of one only component plus the $N_{\rm H}$. By leaving both parameters of the power law free to vary between the different spectra, imposing the NS temperature to be the same, we get $\chi^2_{\nu}=1.25$ (191 dof), while by leaving free the NS atmosphere temperature, imposing the power law (both parameters) to be the same, we get $\chi^2_{\nu}=1.52$ (193 dof). Then, we left the power law (both $\Gamma$ and the normalization) and the NS atmosphere temperature free to vary, together with the $N_{\rm H}$, and we get a better fit,  $\chi^2_{\nu}=1.00$ for 189 dof. The latter model is statistically significant at $5.4\sigma$ confidence level compared to  when the power law alone is free to vary (F-test). We conclude that both spectral components must change, matching the conclusions of \cite{cackett10}.

\begin{figure*}
\begin{center}
\includegraphics[angle=-90,width=5.0in]{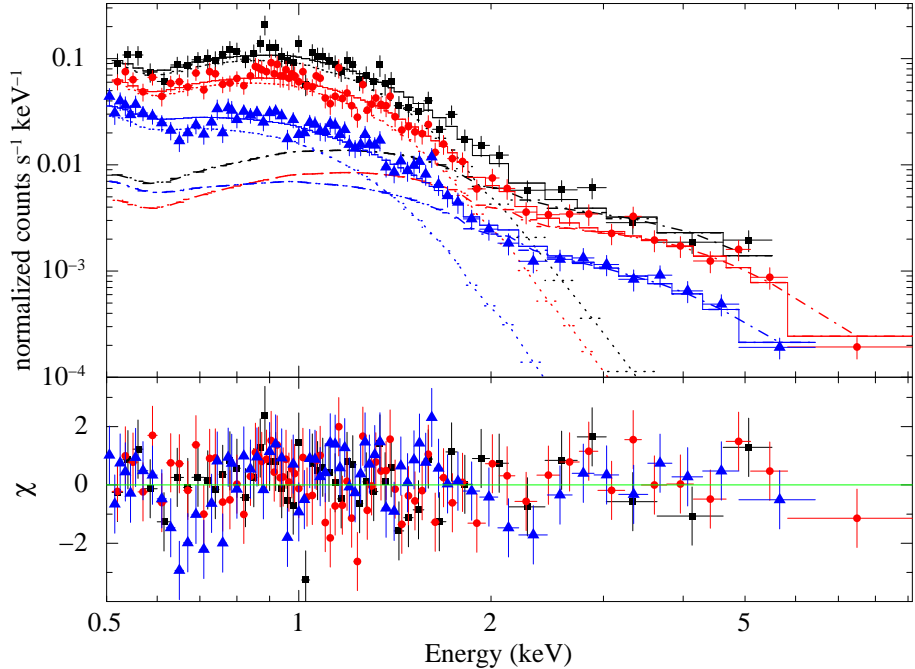} \\ 
\caption{Summed spectra of Cen X-4 for three count rate ranges: low $<0.07$ c/s (blue triangles), medium 0.07--0.11 c/s (red circles), high $>0.11$ c/s (black squares). The solid line represents the total model, while the dotted line represents the \textit{nsatmos} component and the dash dotted line the power law component. Model residuals are showed in the lower panel.}
\label{fig:spec}
\end{center}
\end{figure*}

By leaving both components free to vary, we measure $N_{\rm H}=8.0\pm0.8\times10^{20}$ cm$^{-2}$, which is slightly lower than the total Galactic column density in the source direction \citep[$8.4-9.2\times10^{20}$ cm$^{-2}$]{kalberla05,dickey90}, and slightly higher than that found by \cite{cackett10}, however consistent within cross calibration uncertainty between different satellites \citep[e.g.,][]{tsuji11}. The effective NS atmosphere temperature measured at infinity was found to decrease with the flux: $kT_{h}=80\pm2$ eV, $kT_{m}=73.4\pm0.9$ eV, $kT_{l}=59\pm1.5$ eV. We found a slightly higher value for the power law spectral index in the case of the low count spectrum, where $\Gamma_{l}=2.0\pm0.2$. However, within the $3\sigma$ of the statistical uncertainty it is consistent to be constant, since $\Gamma_{m}=1.4\pm0.2$ and $\Gamma_{h}=1.4\pm0.5$. Only the temperature of the thermal component and the power law normalization are clearly changing (see Tab. \ref{tab:spec}). Since we are mapping here the Wein tail of the thermal component (that peaks at $\sim0.18$ keV), this is equivalent with saying that, within statistical uncertainty, the overall spectral shape is constant and only the flux is changing, meaning that the $nsatmos$ model is moving only up and down in the plot (see Fig. \ref{fig:spec}). Indeed, by imposing that $\Gamma$ is constant between the three spectra, but free to vary, we get a statistically acceptable fit ($\chi^2_{\nu}=1.03$, 191 dof), with $\Gamma=1.7\pm0.15$, and all other model components consistent with the previous fit within statistical uncertainty.

\begin{figure}
\begin{center}
\includegraphics[angle=-90,width=3.0in]{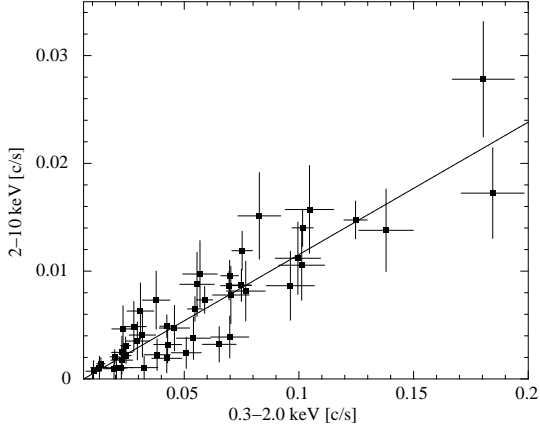} 
\caption{0.3--2.0 keV count rate vs 2--10 keV count rate. The solid line represents the best straight-line fit.}
\label{fig:fluxflux}
\end{center}
\end{figure}

To further assess if the spectral shape is really constant as the flux varies, we compared, for each XRT observation, the 0.3--2 keV count rate with that in the 2--10 keV energy band, where the thermal and power law component are respectively dominating the total flux.  This has the advantage of allowing us to see changes between every daily observation rather than looking at just the three high, medium and low count spectra.  We show this count rate-count rate diagram in Fig. \ref{fig:fluxflux}, and it shows a tight linear relation.  We found that the best fit for the data to be a constant plus a linear function with a constant value of $-8\pm4\times10^{-4}$ and a slope of $0.123\pm0.007$ ($\chi^2_{\nu}=1.14$ for 56 d.o.f.). Since a linear model perfectly fits the data this implies that the overall spectral shape remains constant as the flux changes, meaning that kT is varying and the power law normalization is changing as well and in such a way to compensate the temperature change of the thermal component.

We defined the thermal ratio as the fraction of the unabsorbed 0.5--10 keV thermal flux to the total unabsorbed 0.5--10 keV flux. From the fit with the power law index free we get a thermal ratio of $62\pm6\%$, $59\pm4\%$ and $51\pm5\%$ in the case of high, medium, and low count rate respectively. While, by imposing the power law slope to be the same among different spectra the thermal ratio is more constant and it is $59\pm5\%$, $55\pm5\%$, $54\pm4\%$ respectively. 
We conclude that both component are always responsible for about the same fraction ($\sim50\%$) of the total unabsorbed 0.5--10 keV flux, and they are changing in tandem.  In order to get an estimate of the mass accretion rate, we also extrapolated the spectral model in the 0.01--100 keV energy range, and reported the estimated total flux and luminosity together with that of the thermal component only in Tab.~\ref{tab:spec}.

We note the presence of a feature at about $\sim0.7$ keV in the residual of the low count rate spectrum (see Fig \ref{fig:spec}).  If it is modeled with an absorption Gaussian ($gabs$) we get an energy of $0.68\pm0.01$ keV, a line width ($\sigma$) of $40\pm20$ eV and an optical depth ($\tau$) at line center of $0.08\pm0.02$. The value of $N_{\rm H}$ is slightly lower than in the case of the fit without the absorption Gaussian and it is $6.3\pm0.8\times10^{20}$ cm$^{-2}$, while all other model parameters are consistent within $1\sigma$ of the statistical uncertainty. The fit has a $\chi^2_{\nu}=0.87$ with 186 dof. The inclusion of the Gaussian has a single-trial significance of $4.8\sigma$, however, including the number of spectra and spectral bins searched to find this feature reduces the significance greatly. 
This is the first time that this kind of feature is observed in the spectrum of Cen X-4.
There are several possible lines close to an energy of 0.68 keV \citep[see e.g.][]{verner96}.
In the case that the gas around the NS is in an ADIOS state, a outgoing wind can form driving away matter from the compact object. This is an alternative scenario to the ADAF solution which always involve a radiatively inefficient accretion flow. If such a wind is present, O VIII might be the strongest line detected in the spectrum. 
From the optical depth of the line we estimated the equivalent hydrogen column density due to a possible outgoing wind.
Since $\tau=N\sigma_{c}$, where N is the column density and $\sigma_{c}$ is the Thompson cross section ($6.65\times10^{-25}$), we get N$_{H}=1.20\times10^{23}$  cm$^{-2}$. This is a very large value, a factor of about 150--200 above the photoelectric absorption due to the interstellar medium. Such a dense wind would likely have other observable absorption lines that we do not see, though detailed photoionization modeling would be needed to test this. Given the high density and the lack of this feature in any previous observations of Cen X-4, it is likely not a real detection.

Finally, as a further check of the goodness of the choice of the \textit{nsatmos} model \citep[see also][for a justification of using NS atmosphere models rather than a simple blackbody]{rutledge99}, we verified that the radius $R$, always assumes meaningful values, meaning that it is always consistent with the emission from the NS surface (e.g. is not greatly above 15 km). Consequently, we left $R$ as a free parameter and we estimate its uncertainty at $3\sigma$. We imposed that the atmosphere temperature is the same between different spectra, but left $R$ free to vary and we also imposed $N_{\rm H}=8.0\times10^{20}$ $cm^{-2}$. We obtained an average effective temperature of $66\pm4$ eV and $R_{h}=13\pm4$ km, $R_{m}=12\pm4$ km and $R_{l}=8\pm2$ km ($\chi^2_{\nu}$ 0.96 for 189 dof). We conclude that $R$ is always consistent with the emission from the surface of a NS ($\sim9-15$ km). If instead we fix $R = 10$ km and leave the {\it nsatmos} normalization free (this parameter defines the fraction of the NS surface which is emitting) its value is always consistent with one, within statistical uncertainty.

\begin{table}                                                                                                                                                                                                                                                                                                                                                                                                                                                   
\caption{Main spectral results as a function of the count rate: low $<0.07$ c/s, medium
0.07--0.11 c/s, high $>0.11$ c/s. The three spectra are fitted together, with a total $\chi^2_{\nu}$ for 100 dof.
The measured column density, imposed to be the same 
in the three spectra, is $8.0\pm0.8\times10^{20}$ cm$^{-2}$. 
We assumed a NS Mass of 1.4 M$\odot$, a Radius of 10 km and a distance of 1.2 kpc.
The normalization of the power law is defined as: photons keV$^{-1}$ cm$^{-2}$ s$^{-1}$ at 1 keV. 
We also reported: the thermal fraction, the 0.5--10 keV unabsorbed flux and 
the extrapolated 0.01--100 keV unabsorbed flux, and the derived 0.01--100 keV luminosity at 1.2 kpc, all for both components summed together, and the 0.01--100 keV unabsorbed (bolometric) flux for the thermal component only. Uncertainties are $1\sigma$ confidence level.}
\begin{center}
\begin{tabular}{ccccc}
\hline 
\\
Parameter              & Unity      & High             & Medium         & Low         \\
\hline
$kT^{\infty}$          & eV         & $80\pm2.0$       & $73.4\pm0.9$   & $59\pm1.5$  \\
$\Gamma$               &            & $1.4\pm0.5$      & $1.4\pm0.2$    & $2.0\pm0.2$ \\
Norm.                  & 10$^{-4}$  & $2\pm^{1.7}_{1.0}$ & $1.5\pm0.4$    & $1.3\pm0.3$ \\
Thermal ratio          & \%         & $62\pm6$         & $59\pm4$       & $51\pm5$    \\
&&&&\\
F$^{total}_{0.5-10\, \rm keV}$   &  10$^{-12}$       & $5.0\pm0.5$    & $3.5\pm0.2$ & $1.3\pm0.2$ \\
                               &  erg/s/cm$^2$     &                &             &   \\
F$^{total}_{0.01-100\, \rm keV}$ &  10$^{-12}$ & $\sim14$         & $\sim11$       & $\sim3.5$    \\
                               &  erg/s/cm$^2$ &               &                  &   \\
F$^{thermal}_{0.01-100\, \rm keV}$& 10$^{-12}$  & $5.2\pm0.4$            & $3.7\pm0.2$          & $1.5\pm0.1$       \\ 
                                & erg/s/cm$^2$    &                  &                &              \\                       
L$^{total,\,\rm 1.2\,kpc}_{0.01-100\, \rm keV}$& 10$^{33}$& $\sim2.4$        & $\sim1.9$      & $\sim0.6$   \\
                       & erg/s      &                  &                &              \\
L$^{thermal,\,\rm 1.2\,kpc}_{0.01-100\, \rm keV}$& 10$^{32}$& $9.0\pm0.7$        & $6.4\pm0.7$      & $2.6\pm0.2$   \\
                       & erg/s      &                  &                &              \\

$\chi^2_{\nu}$ (dof)   & \multicolumn{3}{c}{1.00 (100)}                               \\ 
\hline 
\end{tabular}
\label{tab:spec}
\end{center}
\end{table}

\subsection{Spectral Energy Distribution and energetics}
\label{sub:sed}

\begin{figure*}
\begin{center}
\begin{tabular}{cc}
\includegraphics[angle=0,width=3.0in]{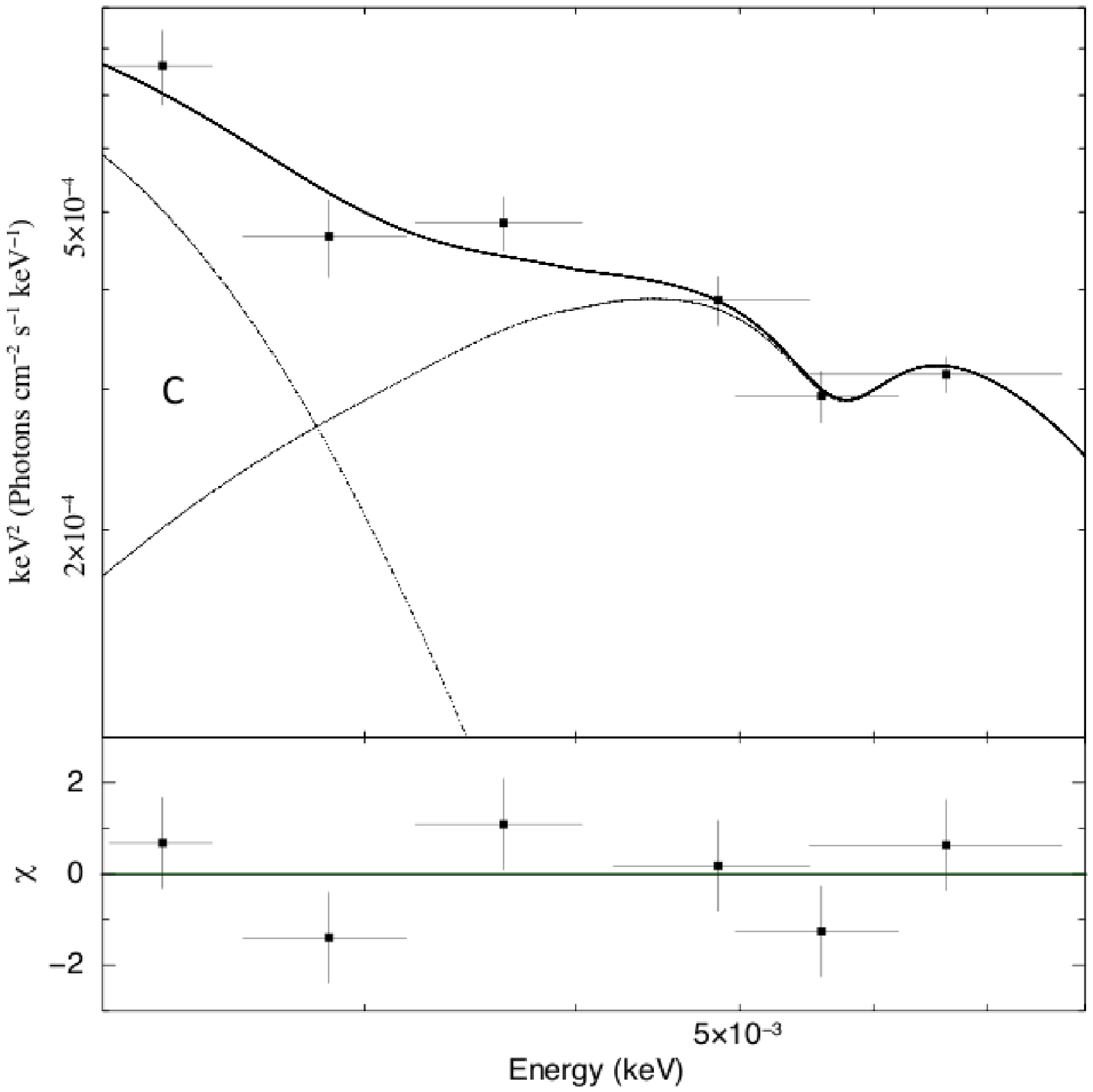} &
\includegraphics[angle=0,width=3.0in]{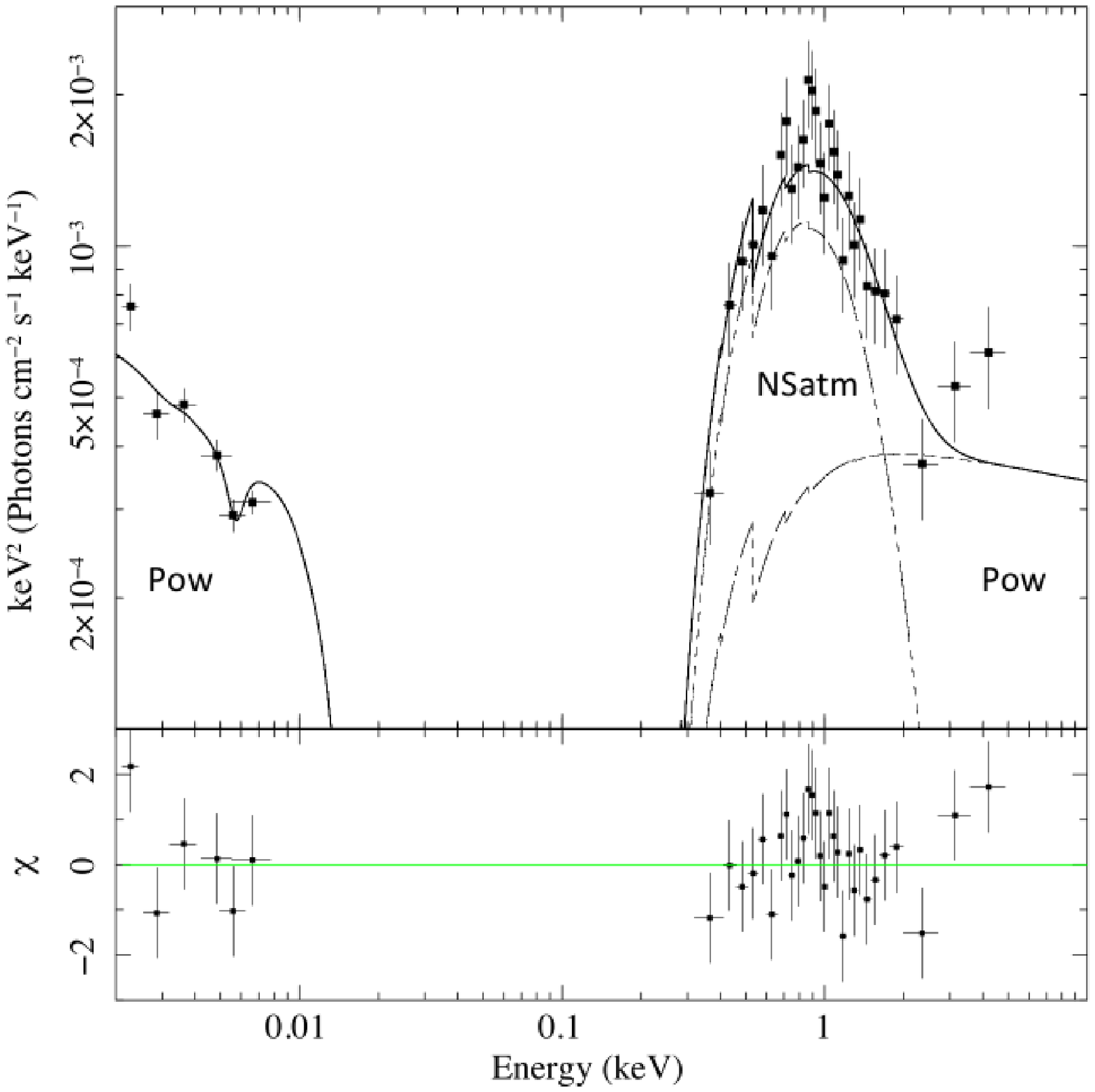} \\
\end{tabular}
\caption{{\it Left panels:} Two blackbody spectral fit in the optical and UV energy range. Dotted lines represent the two blackbody components (that of the companion star, on the left of the plot, is marked with a C), while the continuum line represents the total model. On the lower panel model residuals are shown. {\it Right panels:} Spectral energy distribution (SED) fit from optical up to the X-ray band. A fit with a model comprised of a NS atmosphere ($nsatm$), accounting for the soft X-ray photons ($E<2$ keV), plus a power law (Pow), extending from the optical up to the hard X-ray band and dominating in the optical and UV bands and in the hard X-ray band ($E>2$ keV), is shown. In the case of an ADAF model, the synchrotron emission dominates up to 0.01 keV (substituting the low energy tail of the power law) while the bremsstrahlung dominates above 2 keV (substituting the hard energy tail of the power law).}
\label{fig:2bb}
\end{center}
\end{figure*}

We studied the spectral energy distribution (SED) from the optical to the X-ray band. We have 10 pointings in which \Sw\ was observing in all the UVOT band, so we selected the X-ray brightest (obsid 00035324019, 0.137 c/s) of this 10 pointings in order to get the best S/N possible  (we cross checked the results with that of another bright pointing among these ten, obsid 00035324050, 0.080 c/s). We fixed the reddening at E(B-V)$=0.1$ \citep{chevalier89} using the {\it redden} model in $Xspec$.
We downloaded the correct response matrix file for each filter from the {\it Swift} web page\footnote{{$\rm swift.gsfc.nasa.gov$}}.
We first fitted the UVOT data alone testing the scenario of a companion star (optical) plus an accretion disc (UV). We used a model comprised of two blackbodies (BB), one accounting for the companion star, $c$, at the fixed temperature of $3.88\times10^{-4}$ keV \citep[4500 K]{gonzalez05}, and another one accounting for the disc, $d$, with both its parameters free. Then, we tested a model made by the sum of a BB (accounting for the companion star) with the temperature fixed plus a disc-black body with all component free.  For the two BB model we get: $R_{c}=0.50\pm0.05R_{\odot}$, $kT_{d}=0.0014\pm0.0001$ keV ($16000\pm1400$ K), and $R_{d}=24200\pm200$ km, assuming a distance of 1.2 kpc, with a $\chi^2_{\nu}=1.87$ for 3 dof (see Fig. \ref{fig:2bb}).  For the BB plus disc-BB model we get: $R_{c}=0.43\pm0.06R_{\odot}$, $kT^{inn}_{disc}=0.0024\pm0.0004$ keV ($28000\pm4500$ K), and $R^{inn}_{disc}=8000\pm3000$ km (assuming a disc an inclination of $30^{\circ}$ deg), where kT$^{inn}$ is the temperature of the inner disc and R$^{inn}$ is its realistic radius \citep[see][ for the correction factor between the apparent inner disc radius and the realistic radius]{kubota98}, with a $\chi^2_{\nu}=1.75$ for 3 dof. We note that for both models, the fit is insensitive to kT$_{c}$ if it is left free to vary,  moreover, the $\chi^2_{\nu}$ is not satisfactory.  However, the majority of the residual contribution is arising from the two optical points, where the companion star is dominant. This is not surprising since, here, we are modeling the K3--7 V star with an extremely simplified model, consisting of a pure BB emission. We note that a companion star with a radius of $1.2R_{\odot}$ is inconsistent with our results, which would instead favour a value of $0.5R_{\odot}$, fully compatible with the lower estimate of \cite{gonzalez05}.

We also fitted the whole SED, from the O to the X-ray band. 
In order to first test the scenario proposed by \cite{cam&stell} of a quenched radio pulsar emission, which would result in a single power law extending from the optical up to the X-ray band, we started fitting the SED using a single power law plus a NS atmosphere ({\it nsatmos}) model (to account for the NS surface thermal emission), all multiplied by the reddening ({\it redden}) and the interstellar absorption ({\it phabs}). We fixed the value of following parameters: N$_{H}=8\times10^{20}$ cm$^{-2}$ (average X-ray spectral value); E(B-V)$=0.1$ (known reddening); $R_{NS}=10$ km, $M_{NS}=1.4M_{\odot}$, $d=1.2$ kpc (see Sect. \ref{subs:spec}). The power law (left free to vary) has a photon index $\Gamma=2.10\pm0.03$, and intercepts the optical, UV and hard X-ray points (see Fig. \ref{fig:2bb}). We note that the inclusion of a black body component to account for the companion star is not statistically required, because the power law dominates in the optical band. This is due to the fact that we only have two spectral points in this band. However, a $3\sigma$ upper limit to the radius of the companion star is $0.4R_{\odot}$ \citep[close to the lover estimate of $0.5R_{\odot}$ reported by][]{davanzo05}. 
The model provides a statistically acceptable fit, $\chi^2_{\nu}=0.95$ (33 dof).

Secondly, conscious of the limited number of data points in the UV band, we only tried to put some constrain on the ADAF scenario using a model made by the sum of a Comptonized BB emission ({\it compbb}), plus synchrotron emission ({\it srcut}), plus a bremsstrahlung ({\it bremss}), plus a NS atmosphere ({\it nsatmos}), all multiplied by the reddening ({\it redden}) and the interstellar absorption ({\it phabs}). Comptonization, synchrotron and bremsstrahlung are the three most efficient cooling mechanisms expected in an ADAF for a NS LMXB. The Comptonization is expected to dominate close to the NS while the bremsstrahlung is dominating further out, both mechanisms produce X-ray photons. The contribution of the synchrotron is relevant close to the NS and is emitting in the UV \citep{narayan2}. We fixed the value of the following parameters: N$_{H}=0.08\times10^{20}$ cm$^{-2}$ (average X-ray spectral value); E(B-V)$=0.1$ (known reddening); $kT_{seed}=0.08$ keV (NS surface temperature from spectral fit) and $kT_{e}=86$ keV, $norm=1\times10^{10}$ \citep[as expected for a hot electron Comptonizing medium close to the NS surface emitting at $L=2.4\times10^{33}$ erg/s, for a distance of 1.2 kpc, see][]{narayan2}, where $kT_{seed}$ is the temperature of the seed photons, $kT_{e}$ is the temperature of the hot electrons and $norm=(L_{39}/D_{10})^{1/2}$ is the normalization \footnote{$L_{39}$ is the luminosity in unity of $10^{39}$ erg/s, and $D_{10}$ is the distance in unity of 10 kpc.};$kT_{bremms}=8.6$ keV \citep[as expected for cooler electrons far from the NS at a distance of  $\sim6200$ Schwarzschild radii,][and see also sect. \ref{subs:corrad} for a discussion of this number]{narayan2}.  We found the fit to be insensitive to $\alpha$, which is the spectral index of the synchrotron emission, that consequently we fixed to 1. We also found the break frequency of the synchrotron to be largely uncertain, a lower limit on it being $1.5\times10^{15}$ Hz. This is not surprising since the shape of the observed SED does not show any break in the observed bands, and the flux slowly decreases with the frequency. Consequently, we arbitrarily fixed the break frequency at $2\times10^{15}$ Hz, just above the limit of the explored frequency (see Tab. \ref{tab:ouvflux}). Moreover, the optical depth ($\tau$) of the Comptonizing plasma is also unconstrained. This is because the contribution of this component is minimal both in the optical and UV, where the synchrotron is dominating, than in the hard X-ray (E$>2$ keV), where the bremsstrahlung is dominating (see Fig. \ref{fig:2bb}). Consequently, we removed this component from the total model. The resulting model (NS atmosphere+synchrotron+bremsstrahlung) provides a good fit with $\chi^2_{\nu}=0.97$ with 33 dof. We note that, also in this case, the presence of a black body to account for the companion star is not statistically required, indeed the synchrotron emission dominates at optical wavelength. The $3\sigma$ upper limit to the radius of the companion star is $0.3R_{\odot}$. 


In order to estimate the total flux in the O and UV bands, we convert the flux densities (erg s$^{-1}$ cm$^{-2}$ \AA$^{-1}$), that are the output of {\it uvotsource}, to fluxes by multiplying by the FWHM of the filter bandpass. We use: 769 \AA\ (V), 975 \AA\ (B), 785 \AA\ (U), 693 \AA\ (UVW1), 498 \AA\ (UVM2), 657 \AA\ UVW2 \citep{poole08}. Single flux estimate are reported in Tab. \ref{tab:ouvflux}. We deredden the optical and UV fluxes using the gas-to-dust ratio, $N_H (cm^{-2})=6.86\times10^{21}$ E(B-V), from \cite{guver09} in order to convert from equivalent hydrogen column to E(B-V). Then we use the interstellar extinction curve of \cite{cardelli89} to estimate the extinction correction at the wavelengths of each filter. Consequently, we get the following conversion factors for each filter: 1.396 (V), 1.552 (B), 1.702 (U), 2.054 (UVW1), 2.677 (UVM2), 2.432 (UVW2).

We obtain a UV (except UVM2) flux of $1.06\pm0.04\times10^{-12}$ \ergscm, an optical (B+V) flux of $0.50\pm0.04\times10^{-12}$ \ergscm, and, consequently, a total UV plus optical flux of $1.56\pm0.06\times10^{-12}$ \ergscm. We did not include the UVM2 filter contribution, since it completely overlaps with  the UVW2 and UVW1 band. In order to compare simultaneous O, UV and X-ray flux measure, we also estimated the total unabsorbed X-ray flux. Since the hot NS surface is expected to emit below 0.5 keV, while the power law is likely extending above 10 keV, we extrapolated our model to a wider energy range corresponding to 0.01--100 keV.  Since most part of the 10--100 keV flux is due to the power law component, a small change in its slope can consistently affects the estimated flux.   Consequently, we fixed $\Gamma=1.4$, that is the value obtained on the average high count rate spectrum (and this observation is part of it).
If we consider the uncertainty on the measure of the power law slope,  $\Delta\Gamma=0.5$, we get an X-ray flux ranging from $\sim4.6\times10^{-12}$ \ergscm up to $\sim5.1\times10^{-11}$ \ergscm. Taking into account the uncertainty on the X-ray flux estimate we get that the ratio between UV and X-ray flux has to be $2.1\%<F_{UV}/F_{X}<23\%$.

\begin{table}
\caption{Estimated flux in the O-UV filter bandpass for obsid 00035324019. The corresponding 
$\nu F_{\nu}$ value is also reported. Uncertainties are $1\sigma$ confidence level.}
\begin{center}
\begin{tabular}{cccc}
\hline 
\\
filter & $\nu$      & Flux               & $\nu F_{\nu}$       \\
       & $10^{14}$  & $10^{-13}$         & $10^{-13}$  \\
       & Hz         &  \ergscm           &   \ergscm   \\
\hline  
UVW2   & 15.5       & $4.4\pm0.2$        &  $5.3\pm0.2$       \\
UVM2   & 13.3       & $2.9\pm0.2$        &  $4.9\pm0.3$       \\
UVW1   & 11.5       & $3.2\pm0.3$        &  $6.0\pm0.4$       \\
U      & 8.7        & $3.0\pm0.2$        &  $7.9\pm0.7$       \\
B      & 6.8        & $2.6\pm0.3$        &  $7.6\pm0.7$       \\
V      & 5.5        & $2.4\pm0.2$        &  $12\pm1$          \\
\hline              
\end{tabular}
\label{tab:ouvflux}
\end{center}
\end{table}

\section{Discussion}
\label{sec:discuss}

\subsection{Evidence for accretion}

We showed that the X-ray and UV light-curve of Cen X-4 during quiescence are both highly variable, with the fractional root mean square variability F$_{\rm var} = 73.0\pm1.5\%$ and $50.0\pm1.4\%$ respectively, while the optical light curve is changing less (F$_{\rm var}=10.0\pm1.6\%$). The first order X-ray and UVW1 structure function ($V_{\tau}$) shows the shape expected for a power spectrum with index -2 (red noise), as is typically seen in accreting systems (two plateau connected by a power law). The timescale of the variability underlying the structure function goes from few days ($\sim5$) up to months. The excess in the structure function at about 5 days is likely connected to the peaks which are present in the X-ray and UVW1 light curves at about 40, 50, 70, 90, and 110 days, whose average rise and decay time is indeed about 5 days. In the time between different peaks the count rate seems to always reach a minimum (ground) level extremely close to the detection limit for both the X-ray and the UVW1 band. We observed the strongest X-ray variability on short timescale ever detected for Cen X-4 during quiescence\footnote{We are not considering here X-ray bursts detected from Cen X-4 immediately before outbursts began or the X-ray flare in 1971 detected by Apollo 15 that could have been an X-ray burst from Cen X-4 \citep{kuulk09}. The association with Cen X-4 is not certain and is impossible to establish if the source was truly in quiescence at that time.} a factor of 22 and a factor of 13 in only four days (between 97.2 d and 101.5 d and between 39.5 d and 43.3 d respectively). This is much higher than previously reported by \cite{campana97}, who found a factor of $\sim3$ on a timescale of four days and \citet{cackett13} who found a factor of $\sim3$ change in 7 days.

Moreover, we showed that both the spectral components, the NS atmosphere and the power law, must change with the flux.  We also showed that they are changing in tandem, implying that they are closely linked. While this has previously been shown for timescales of months to years \citep{cackett10}, this is the first demonstration on timescales of days. A change in the thermal component that we detected as a surface temperature change strongly suggests that the accreting matter must reach the surface of the NS, overtaking the centrifugal barrier of the rotating magnetosphere. Since the two spectral component are so closely linked, accretion must be, very likely, also responsible for the change in the power law. A very low level of accretion seems to occur quite continuously, generating random episodes of high count rates and consequently modifying the temperature of the surface, which is hotter when the count rate is higher. We emphasize that, assuming the NS atmosphere model to be the correct model to fit the data, we obtain a radius which is less than 15 km. This is encouraging in considering the X-ray emission as arising from the NS itself. The scenario proposed by e.g. \cite{cam&stell}, where the X-ray are produced at the shock point between a radio pulsar wind and the inflowing matter from the donor star, is for Cen X-4 likely disfavoured. While this scenario can account for the single power law extending from the O-UV up to the X-ray band, according to its prediction only the power law component should change, which is not the case. We note that it is extremely difficult to account for such intense, non monotonic, multi-flare-like variability, with a clear correlation between the thermal and power law spectral component, without invoking accretion that is somehow occurring at very low Eddington luminosity rates in Cen X-4. However, we emphasize that for other quiescent sources, where e.g. the thermal and power law emission are not so closely correlated, accretion may not be the right scenario.  

When matter is accreted onto the NS surface it can emit a thermal-like spectrum \citep{zampieri95}, that is harder than a BB emission at the NS effective temperature, moreover, it also emits a consistent excess in the Wien tail region. At the level of S/N of our data, this perfectly matches the characteristics of a NS atmosphere model \citep[see][for more details about the comparison between different thermal models]{soria11}. Cen X-4 is not the only NS LMXB showing quiescent variability other example are Aql~X-1 and EXO~1745$-$248 \citep[see e.g.][and references therein]{cackett11, natalie12}, but at present it is the source showing the most intense variation on short timescales (4 days). More recently, SAX~J1750.8-2900 showed evidence of a small flare, with flux increase of a factor of 10 above the quiescent level \citep{rudy13}, lasting less than 16 d, suggesting that also in this case low level of accretion can also occur in quiescence at least during these flare episodes \citep[see also][for a similar flare from the transient NS XTE J1701-462]{fridriksson11b}. This kind of flare is reminiscent of the peaks in the count rate we are observing from Cen X-4. They both last several days, however Cen X-4 has a less intense emission, $L=10^{32-33}$ erg/s (compared to $10^{34}-4\times10^{34}$ erg/s). We can detect and explore this kind of variability, occurring at low X-ray luminosity, only for Cen X-4, because it is relatively close to us.

\subsection{Remarks about past and future quiescent studies}


We emphasize that, because of the strong intensity change in the UV and X-ray count rate we observed here, any kind of study of the quiescent emission of Cen X-4, and of LMXB in general, must be based on simultaneous multiwavelength (e.g. O, UV, and X-ray) data only.  Recently, \cite{hynes12} pointed out that a promising method to discriminate between BH and NS LMXB is to compare their X-ray vs UV luminosity ratio, with the NS exhibiting $L_{X}/L_{UV}$ a factor of 10 times higher. However, only two sources in this study had simultaneous X-ray and UV observations \citep[note that][do caution about the use of non-simultaneous data]{hynes12}. We showed that using non-simultaneous data can easily introduce an error of at least a factor $\sim22$ in the estimate of the ratio. The same kind of consideration and caution also applies to any other multiwavelength quiescent study where non-simultaneous data are used.

If accretion is still occurring also at low Eddington luminosity rates for Cen X-4, this would imply that any intrinsic cooling emission from the NS should be equal or less than the observed kT in the lowest count rate spectrum, $59.0\pm1.5$ eV. In cooling NS studies the quiescent  bolometric thermal luminosity (e.g. 0.01-100 keV), that for Cen X-4 we measured to be $\leqslant2.6\times10^{32}$ erg/s, can be compared with the estimated time-averaged mass transfer rate to find the most likely cooling scenarios among the several that are currently available \citep[see e.g.][and more references therein]{heinke09,wijnands13}. The literature value of the surface temperature of Cen X-4 used for these kind of studies, $kT=0.76$ \citep{heinke07}, is higher that we measure here for the low count rate spectrum, consequently also the inferred luminosity is higher $4.8\times10^{32}$ erg/s. Moreover, we also note that the time-averaged mass transfer rate is derived under the condition that the time-averaged mass accretion rate over the last 10 years is reflecting the time-averaged mass transfer rate. However, we have shown that the emission from Cen X-4 in quiescence is highly variable, with count rate variation of at least a factor of 22 in only few days. Summarizing, we emphasize that to perform an accurate study of the cooling mechanism of Cen X-4 a temperature $\leqslant59$ eV, and an 0.01--100 keV bolometric thermal luminosity $\leqslant2.6\times10^{32}$ erg/s, should be considered.

\subsection{Magnetospheric accretion without a strong propeller?}

Models of accretion flows around quiescent NSs like \mbox{Cen X-4}
typically assume the source is in a ``strong propeller'' regime
\citep{1975A&A....39..185I}, in which the rapidly spinning
magnetosphere of the star creates a large centrifugal barrier that
drives an outflow and inhibits accretion onto the star. The persistent
X-ray emission and rapid variability seen in \mbox{Cen X-4} and other sources
challenge this assumption.

Furthermore, \cite{asai98,1999ApJ...520..276M} 
have suggested that a powerful propeller effect is necessary to make
an ADAF model compatible with observations of quiescent neutron
stars. This is because ADAFs radiate very inefficiently but accrete at
a relatively high rate even in quiescence ($\sim 10^{-2}-10^{-3}\dot{M}_{\rm Edd}$). 
The ADAF picture should apply to both quiescent
NSs and BHs, except that NSs will have a
much higher radiative efficiency than BHs, since all the
matter advected through the flow will end up on the surface of the
star. To reconcile this fact with the low quiescent luminosities of
NSs like \mbox{Cen X-4} ($\sim 10^{-5} L_{\rm Edd}$),
\cite{1999ApJ...520..276M} have suggested that the vast majority (up
to $\simeq 99.9\%$) of the accretion flow around the star is expelled
through the propeller effect.

However, \mbox{Cen X-4} likely has a low magnetic field, 
indicating that a strong propeller is unlikely to be operating 
at the accretion rates implied by the ADAF scenario. 
In order for matter to be expelled in a propeller, the ``magnetospheric radius'' (where the
magnetic field is strong enough to truncate the disc) must
significantly exceed the corotation radius ($r_{\rm c} \equiv
(GM_*/\Omega^2_*)^{1/3}$, where $M_*$ is the mass of the star and
$\Omega_*$ its spin frequency), the radius at which the Keplerian
rotation in the disc equals the spin of the star
\cite{1993ApJ...402..593S}. Here we assume (favourably for a propeller to
form) that \mbox{Cen X-4} has a spin period of $\sim 2.5~\rm{ms}$ and
the magnetic field is of order $10^9$G (still small enough to be
overwhelmed by $\dot{M}_{\rm Edd}$ so that pulsations are not seen in
outburst; in reality fields $\sim10^7-10^8~{\rm G}$ are more
likely). If $\dot{M} \simeq 10^{-2} \dot{M}_{\rm Edd}$ through the
flow as it approaches the star (with 0.1\% of that eventually
accreting onto the star), the accretion flow will be truncated around:
\begin{equation}
r_{\rm m} = 6.3 R_*\frac{\xi}{0.52}\left(\frac{B}{10^9
  \rm{G}}\right)^{4/7}\left(\frac{0.01}{\dot{M_{\rm
      Edd}}}\right)^{-2/7}\left(\frac{M_*}{1.4M_\odot}\right)^{-1/7}.
\end{equation}
In this equation \citep[taken from][]{ghosh79}, $R_*$ is
the NS radius, $B$ is the surface magnetic field and $\xi$
is a factor of order unity to account for uncertainties in the
interaction between the flow and stellar magnetic field. Therefore,
even under favourable circumstances, $r_{\rm m} \sim 2 r_{\rm c}$,
which is unlikely to produce such a strong propeller as needed in this
picture. Recent simulations of the propeller regime find $\dot{M}_{\rm
  acc}/\dot{M}_{\rm ej} \sim 0.04$ (the ratio of accreted to ejected
material) for a considerably stronger propeller ($r_{\rm m} \sim 5
r_{\rm c}$; \citealt{2013arXiv1304.2703L}).  Moreover, the ratio
between accreted and ejected matter (and how it scales with $r_{\rm
  m}/r_{\rm c}$) is extremely uncertain, making it essentially
impossible to predict the mean accretion rate through the flow from
the amount that ends up on the star.

The extremely low accretion/ejection ratio required by the
propeller-ADAF model also presents a challenge for mass-loss from the
donor star. We estimated that over the last 40 years \mbox{Cen X-4} accreted $\sim
(4-7)\times10^{24}~{\rm g}$ during its two outbursts and $\sim
10^{25}~{\rm g} $ during quiescence (assuming $\dot{M}_{\rm
  acc}/\dot{M}_{\rm ej} \sim 10^{-3}$, i.e. the outflow rate is
1000$\times$ larger than the observed accretion rate). This implies a
mass transfer rate of $2\times 10^{16} \rm{g}~\rm{s}^{-1}$. Without an
outflow, the transfer rate is $(3-6)\times
10^{15}\rm{g}~\rm{s}^{-1}$. Models for mass transfer through
Roche-lobe overflow suggest transfer rates of at most $(2-7)\times
10^{15} \rm{g}~\rm{s}^{-1}$ (\citealt{1999ApJ...520..276M}; although
they note enhanced accretion from magnetic braking is possible).

An alternative scenario (`dead disk') to the strong propeller was recently presented
in \cite{2010MNRAS.406.1208D,2012MNRAS.420..416D}. Following from an
earlier suggestion of \cite{1993ApJ...402..593S}, they note that for
$r_{\rm m} < 1.3 r_{\rm c}$ (which could likely be the case of Cen X-4) 
the interaction with the magnetic field
will not be enough to unbind the disc into an outflow, so that
material at $r_{\rm m}$ will instead remain bound, adding the angular
momentum it acquired to the accretion flow. 
More significantly, in this scenario the inner edge
of the accretion flow remains always very close to $r_{\rm c}$, 
and thus allows accretion onto the star, even at very low accretion rates. 
Meaning that the magnetosphere holds back material in the flow, but some small amount accretes onto the star.
This may naturally explain the low-level quiescent emission seen in
Cen X-4, which could then be a reflection of $\dot{M} \simeq 10^{-5}-10^{-6} \dot{M}_{\rm Edd}$ 
in the inner regions of the flow. 
This can happen in principle both for an ADAF and for a thin disc scenario, however,
a note of caution: the solutions of 
\cite{2010MNRAS.406.1208D} were formally calculated assuming a thin accretion disc, 
and may be significantly altered for an ADAF-type solution.

In order for the scenario outlined by \cite{menou99} 
to still be viable, the vast majority (99.9\%) of the mass 
in an ADAF must be expelled before it reaches the magnetosphere of the NS. 
This must then apply equally to ADAFs around NSs and BHs, 
and support models \citep[see e.g.][about the ADIOS model]{blandford99} 
in which most matter in a radiatively-inefficient 
flow is expelled before it accretes onto the central 
object by an intense outgoing wind intrinsic to the accretion flow
itself.

\subsection{Origin of the UV emission}

In quiescent LMXBs the donor stars are considered intrinsically too cool to have any significant UV emission \citep[e.g.,][]{hynes12}. The UV, then, could probe the emission from a particular region of the accretion flow. However, the exact location of origin of the UV emission in the accretion flow is still unclear and it is consequently debated. The UV emission could arise due to:  (I) the thermal emission from the gas stream-impact point,  (II) the thermal emission from a standard, optically thick and geometrically thin accretion disc truncated far from the compact object, (III) the emission from an advection dominated accretion flow (ADAF), or alternatively the UV emission could be due to (IV) X-ray irradiation of, and reprocessing (perhaps in the inner accretion disc or on the surface of the companion star). We will explore in detail the different scenarios in the following section.

\subsubsection{Thermal emission from the gas stream-impact point}

Our analysis clearly showed that the UV and the X-ray emission are strongly correlated. Signs of a correlation were first noticed by \citep{cackett13}. However, thanks to our long term multi-wavelength monitoring campaign, we had the chance to deeply explore the nature of this correlation. We find that X-ray and UV emission are correlated on long timescale, t$\leq5$ ks (corresponding to an XRT pointing), on short timescale, t$\leq2$ ks (corresponding to an UVOT snapshot), and also on very short timescale down to t$\leq$110 s (corresponding to the shortest UVOT snapshots). Furthermore, we have been able to unveil for the first time the real shape of this correlation. Whenever enough data points are available so that the high X-ray count rate ($\gtrsim0.17$ c/s) is covered, the shape of the correlation is a power law with spectral index 0.2--0.6. Moreover, also the optical V band and X-ray emission are correlated, both on long and on short timescales. The presence of a correlation, especially that at very short timescales (110 s) clearly rules out the thermal emission from the stream impact point as source of the UV photons. Indeed, if we assume that the UV light is tracing the matter in the disc, UV variations would refer to instability propagation throughout the disc. However, the timescale to traverse the whole disc from the stream impact point (where the matter lost by the companion star impact on the outer accretion disc) to the inner edge is significantly longer than the correlation timescale, being of the order of several weeks \citep{fkr92}. Such a long timescale cannot explain any of the correlations we are observing here. These results are in contrast with what reported by \cite{mccli00} and \cite{menou01}, who found the gas stream impact point as the most likely source of the UV emission for Cen X-4. However, they only suggested that the UV emission could be powered, in term of total luminosity, by the estimated mass transferred by the companion star to the outer edge of the accretion disc. Nevertheless, they could not explore the consequences of variability, crucial to test this scenario, due to the lack of data at that time. 
In summary, the fact that the X-ray and UV emission are correlated unambiguously rules out the stream impact point as source of the UV photons.

\subsubsection{Thermal emission from a truncated standard accretion disc}
\label{subsub:thermal}
No sign of the presence of an accretion disc has been seen in the X-ray spectrum of Cen X-4 during quiescence, or in LMXBs in general -- the thermal component in quiescent NSs is always consistent with the NS itself, while quiescent BH just show power-law spectra. This suggests that, at very low mass accretion rate, the disc is likely not extending down to the NS surface, as occurs during outburst (where the disk is clearly emitting in the X-ray), but is likely truncated further out. 
Theoretical model gives support to the truncated disc scenario, suggesting that is extremely easy for the matter in the inner part of the accretion disc, if the mass accretion rate is very low (e.g. under a critical value that for NS is $\dot{M}_{cr}/\dot{M}_{Edd}\sim10^{-2}-10^{-3}$), to evaporate and enter in a optically thin and radiatively inefficient spherical accretion flow state surrounding the compact object \citep{narayan1,narayan2,narayanbook,blandford99}. 
Moreover, the DIM model \citep[][]{cannizzo93,laso} require the disk to be truncated during quiescence in order to explain transients.
Based on the correlation results only, we now want to demonstrate that mass accretion rate fluctuations, propagating inwards from any point of a standard thin disc \citep[e.g.]{shakura73}, can not be the reason for the X-ray variability observed here.

In the standard disc solution the radial velocity of the matter in the disc is of the order of 0.3 km s$^{-1}$ \citep{fkr92}, and the measured timescale of the correlation is t$\leq110$ s. Consequently, if the observed correlation is generated by matter propagating in a standard disc, finally accreting on the NS surface, this matter must propagate from a maximum distance of about 33 km from the NS center. However, this would imply that the disc is extending down to the NS surface which is not observed (the X-ray thermal emission is consistent with the emission from the NS surface only).  Moreover, if we impose that the viscous timescale, which is the timescale on which matter moves through the disc under the influence of the viscous torque, is equal to the correlation timescale, we could get an estimate of the size of the disc in an extremely peculiar scenario, where the dynamics in the disc would be fast enough to justify the observed correlation. The radius of the accretion disc as a function of the viscosity timescale is given by Eq. \ref{eq1} \citep{fkr92}.
\beq
\label{eq1}
R_{10}=(\frac{t_{vis}}{3\times10^{5}})^{4/5}\alpha^{16/25}m_{1}^{-1/5}(\dot{M}_{16})^{6/25}
\eeq
where $R_{10}=R/(\rm{10^{10}\,cm})$ is the disc size, measured from the center of the accreting compact object, $\alpha=0.1$ is the viscous parameters, $\dot{M}_{16}=\dot{M}/(\rm{10^{16}\,g\,s^{-1}})$ is the mass accretion rate, and $m_{1}=M/M_{\odot}=1.4$ is the compact object mass. In order to maximize $R_{10}$, we estimate the mass accretion rate from the maximum 0.01--100 keV X-ray luminosity (high count rate), $L_{X}\sim2.4\times10^{33}$ erg s$^{-1}$, and since $\dot{M}=L_{X}/\eta\,c^{2}$ and $\eta=0.1$ we get $\dot{M}_{16}\sim2.7\times10^{-3}$. By imposing the viscous timescale to be equal to the shortest correlation timescale found, we can derive the maximum size of the disc for the specific timescale. By using t$_{c}\leq110$ s (correlation timescale in the U band) we derive $R\lesssim9$ km, a totally non physical value, as it is smaller than the size of the NS itself.  We note that also considering $\alpha=1$ we get R$\lesssim40$ km, which is once again a non physical value, because the disc is not extending so close to the NS surface.
We conclude that a standard disc cannot account for the observed correlation, implying that UV variability does not arise from instabilities propagating through a standard disc.

Finally, we also note that assuming that a BB and/or a disc BB model are reasonable approximations of the real disc spectrum, the temperatures derived from the SED fit for the inner disc edge is of the order of 16000--28000 K. At this temperature the disc would experience an immediate outburst \citep[][and references therein]{menou01}. So, the possibility that the UV is arising from a standard disc is also ruled out, in an independent way, based on the SED results. 

\subsubsection{Advection dominated accretion flow, ADAF}
\label{subs:corrad}

 
According to \cite{narayan1} and \cite{narayan2}, when the mass accretion rate, and the ratio $r$, drop below a critical (cr) value, $r_{cr}=\dot{M}_{cr}/\dot{M}_{Edd}$, which in the case of an accreting NS if $\alpha=0.1$ is $\sim0.1\alpha^{2}\sim10^{-3}$, the accretion flow can only exhibit in two distinct stable states, namely the Shakura \& Sunyaev thin disc and an advection dominated accretion flow (ADAF). In ADAF the accretion flow is radiatively inefficient, and the energy produced by viscosity is advected by the flow and stored as heat, rather than be radiated away. Moreover, \cite{narayan2} also suggested that whenever $\dot{M}<\dot{M}_{cr}$, the accretion flow always selects the ADAF solution, as it is the only really stable one. Independently from this last statement, we have already demonstrated that the standard disc can not be the source of the observed UV variability. Since for 
Cen X-4 $r$ is of the order of $10^{-6}$ we now explore the possibility that the correlation arises from an ADAF region, where UV flux variation could still trigger X-ray variation.

In an ADAF the matter moves towards the disc with a radial velocity $v_{r}\sim\alpha\,c_{s}$. The speed of sound is $c_{s}\sim v_\mathit{ff}$, where $v_\mathit{ff}=(GM/R)^{1/2}$ is the free fall velocity \citep{narayan1}. For a reasonable value of $\alpha$, $\sim0.1$, we have $v_{r}=0.1(GM/R)^{1/2}$. Since the timescale at which accretion occurs is defined as $t_{acc}=R/v_{r}$, where R is the distance from the center of the accreting object, we have:
\begin{equation}
R=(t_{acc}\,0.1)^{2/3}(GM)^{1/3} 
$\label{eq:rcorr}$
\end{equation}
We can estimate the maximum distance to the NS from which the matter is accreting in an ADAF (R$_{acc}$),  assuming that the accretion timescale is equal to the shorter correlation timescale we found ($t_{cor}\leq110$ s). This is a natural assumption since only if $t_{acc}<t_{cor}$ can accretion via an ADAF be the cause of the observed correlation. From Eq. \ref{eq:rcorr}, where $M=1.4$ M$\odot$, we get R$_{acc}\lesssim25000$ km, which corresponds to $\sim6200$ Schwarzschild radii, in agreement with theoretical values \citep{narayan2}. Consequently, the ADAF scenario can account for the timescale of the correlation. The variability in the amount of matter that the star is accreting from a region inside of a radius of $R\sim6200R_{s}$ could trigger X-ray variability. This radius must be considered as the maximum distance from which matter can be accreted on the NS in ADAF, based on the observed correlation. However, with present spectral data, we were unable to verify if  the ADAF model can accurately reproduce the whole source SED (from optical to X-ray). The number of points in the O and UV band is limited to 6, and the coverage is likely not wide enough. In fact, as already noticed by \cite{mccli00}, the peak of the thermal synchrotron emission ({\it srcut} model used in the SED fit) depends on the mass of the compact object, $\nu_{s}\sim10^{15}(M/M_{\odot})^{-1/2}$, and for any reasonable value of Cen X-4 NS mass, $\nu_{s}$ would lie well above the highest inspected frequency being $\nu_{s}>6\times10^{15}$ Hz. However, instrumentally explore higher frequency is not easily possible. On the other hand, \cite{menou01} modeled HST/STIS data of Cen X-4 with an ADAF plus a magnetosphere producing propeller effect. The magnetosphere complicates the scenario, likely truncating the innermost and hottest part of the accretion flow, where the brightest and high energetic (higher frequency) synchrotron emission should be generated. They argued that the synchrotron emission from an ADAF can not account for the intensity and frequency peak of the observed UV emission and, consequently, concluded that for quiescent NS LMXB the contribution of an ADAF to the UV light must be very little. Summarizing, while the ADAF alone is very promising to explain the timescale of the observed accretion, however, it seems to be unable to explain 
the intensity and the peak of the UV emission without invoking some still not well understood extra effect.

\subsubsection{X-ray irradiation and reprocessing}

Another scenario proposed to explain the observed correlation takes into account the irradiation and reprocessing of the X-ray, arising from (or near to) the NS surface, by part of the accretion disc (usually the inner part which will be hottest) as was suggested to explain the X-ray-UV correlation in Cen~X-4 by \citet{cackett13}. We note that the light-travel time from the NS to the outer edge of the disc is only about 5 s (the outer disc radius for a 15.1 h binary with a NS of $1.4M_{\odot}$ is $\sim1.5\times10^{11}$ cm), so perfectly consistent with the reprocessing scenario.

The X-ray reprocessing from accretion disc was studied by \cite{vanpar94} which found that the optical luminosity scales with the X-ray luminosity as $L_{V}\propto\,L_{X}^{1/2}\,R_{out}$, where $R_{out}$ is the radius of the outer accretion disc. This matches our results, as the most likely shape of the correlation is a power law, with index $\gamma$ always consistent, within $3\sigma$ of the statistical uncertainty, with 0.5. 

Let us now explore in more detail the possibility that the X-rays are reprocessed by a standard disc at a radius $R$ from the NS surface. According to the reprocessing scenario the UV flux has to be only a small fraction of the X-ray one. More specifically, the maximum fraction of the reprocessed emission $S$ has to be equal to the fraction ($S^{'}$) of the sphere surface of radius $R$ occupied by that part of the disc reprocessing the X-ray radiation. $S=S^{'}$ only works if the reprocessing is $100\%$ efficient, otherwise $S<S^{'}$. We emphasize that for a standard disc, $S^{'}$ is always extremely small, as the disc is geometrically thin.  In more detail, if $H$ is the vertical extension of the disc, we have $S^{'}\simeq H/2R$. The ratio $H/R$ for a standard disc which is emitting because of irradiation, is given in \cite{fkr92} as:
\beq
H/R=1.7\times10^{-2}\alpha^{-1/10}\dot{M_{16}}^{3/20}m_{1}^{-3/8}R_{10}^{2/7}f^{3/5}
\eeq
where $f=[1-(R_{NS}/R)^{1/2}]^{1/4}$. $H/R$, and consequently $f$, are maximum at the outer disc (where $f\sim1$ and where $R=R_{10}\sim15$, if $P=15.1$ h). Since $\dot{M_{16}}\sim2.7\times10^{-3}$ (see Sec. \ref{subsub:thermal} for the estimate of $\dot{M_{16}}$), this implies that $S^{'}\leq0.8\%$.  We stress that since $H/R\propto\dot{M_{16}}^{3/20}$, even if Cen~X-4 was emitting at the Eddington limit, $S^{'}$ would remain small, $\sim5\%$. However, the value of $S^{'}$ does not match the amount of UV reprocessing estimated from the observation, where $S=F_{UV}/F_{X}$ is $2.1\%<S<23\%$ (see Sec. \ref{sub:sed}). We therefore conclude that $2.1-23\%$ is not a reasonable value for reprocessed emission from a standard thin disc \citep[note that this is in contrast to the conclusions of][who did not fully consider the geometric size of disc]{cackett13}. Our conclusions are even stronger if we take into account only the inner edge of the disc where H/R is minimal. We note that in order to have $F_{UV}/F_{X}\leq0.8\%$, as the observed UV flux is $1.06\times10^{-12}$ \ergscm ($L_{UV}=1.83\times10^{32}$ erg/s), the 0.01--100 keV X-ray flux should be  $F_{X}\geq1.3\times10^{-10}$ \ergscm ($L_{X}\geq2.2\times10^{34}$ erg/s). This is well above the maximum quiescent luminosity ever recorded from the source. 

We emphasize that, however, the real structure of the disc in quiescence is unknown, and it is likely different from the standard disc configuration. For example the value $\alpha$ could depend on the radius R (while the standard disc solution assumes that $\alpha$ is constant), and this can make the scenario more complex. Among other things, the vertical size of the outer disc could be greater than envisaged in simplified treatments of the inner disc. In any case, in order to be the disc alone to reprocess e.g. 5\% of the X-ray emission, its vertical size must be $H\geq10\%R$. Any disc model invoked to account for reprocessing must be able to account for that.

On the contrary, we note that the fraction of the sphere of radius $R$ occupied by the surface of the companion star of radius $R_{c}$ is consistent with the observed fraction of reprocessed light. Indeed, in this case $S^{'}\simeq1/2(R_{c}/R)^{2}$. The companion star is a K3-7 V star with radius $0.5<R_{c}/R_{\odot}<1.2$, and that the mass ratio of Cen X-4 is $q=0.17\pm0.06$. By using the Eggleton's equation we can derive the ratio $R_{c}/R = 0.49q^{2/3}/[0.6q^{2/3}+ln(1+q^{1/3})]$. Taking into account the $1\sigma$ uncertainty on the measure of $q$ we get $2.3\%<S^{'}<3.4\%$, which matches our experimental results. We conclude and stress that the companion star is the most likely source of reprocessing, while the accretion disc alone cannot.

We emphasize that the SED, in addition to the emission from the companion star,  also indicates the presence of a small ($\sim2.4\times10^{4}$ km) hot ($kT\sim0.0014$ keV, $\sim16000$ K), thermally emitting region, that we demonstrated is not the accretion disc.  This could be a hot spot on the star companion, which is likely hotter because directly (without any shield from the accretion disc) irradiated by the X-ray photons. However, a higher number of data points in the UV band (compared to our 4) is mandatory to properly characterize this component. The $3\sigma$ upper limit on the size of the spot is about 43000 km. This corresponds to a fractional surface area of about $2\%$ for a $0.5R_{\odot}$ companion (that is the companion radius estimated from the SED fit). 
We note that this UV component has a small area. However, we used a simple black body model to fit for something which is likely more complex. Moreover, we can speculate that the X-ray reprocessing is likely generating a gradient of temperature over the companion star. Due to the limited number of points (4) in this band we can highlight and measure only where this gradient is maximum, likely a small region closer to the NS.

\cite{torres02} argued that the X-ray irradiating the inner face of the companion star can generate a narrow $H\alpha$ line detected in the optical spectrum. Furthermore, the contribution in the optical of a hot spot on the companion star due to X-ray heating is required to model the optical and IR light-curve of Cen~X-4 \citep{mccli90,shahabaz93}. More recently, accurate Doppler tomography of Cen~X-4  was performed by \cite{davanzo05}. They found strong $H\alpha$ and HeI ($\lambda5876$ \AA) emission from the companion due to irradiation from the NS. Moreover, they even found the outer disc to be irradiated. Indeed the outer disc is hotter than the companion, emitting optical HeI ($\lambda6678$ \AA) lines. They argue that the $H\alpha$ emission from the secondary originated in a region close to the Lagrangian point, where the X-ray radiation is mainly occulted by the disc, and consequently this region is cooler, while the HeI ($\lambda5876$ \AA) might arise from the portion of the companion star surface which is directly irradiated by the X-ray (not blocked by the disc), and hence this region is hotter. All these results support the idea of a hot spot on the star due to X-ray irradiation.

We also found a correlation between the X-ray and the optical V band emission.  This was not seen in \citet{cackett13} where they have significantly fewer points, and is only revealed here by the much larger dataset and dynamic range in X-ray flux.  If we compare the constant plus power law model to fit for the shape of the correlation in the case of the V band ($t\leq5$ ks and $t\leq2$ ks), with that of the UV bands, we note that while for a low X-ray count rate the UV count rate goes extremely close to zero, this is not the case for the V band. 
This suggests that there is an intrinsic level of emission in the V band, independently from the physical mechanism causing the correlation, and it is much higher than that in the UV bands. We now estimate, for each band, the ground level of emission and its contribution to the total light. We measure the value of the constant in the fit which is 0.5 c/s in the V band compared to a $3\sigma$ upper limit of $0.02$ c/s in the case of the UVW1 band. We note that 0.5 c/s is exactly the ground level from which the V band lightcurve displays its variability (see Fig. \ref{fig:3b_lc.ps}). 
Taking into account that the maximum V count rate is $\sim0.8$ c/s, this corresponds to a $60\%$ increase from 0.5 c/s. This means that at least $40\%$ of the total emission in the V band, is stable, and likely due to the companion star which, indeed, is expected to dominate at optical wavelengths \citep{chevalier89}. However, part of the intrinsic V emission can also arise from the accretion disc, but with present data we can not quantify it \citep[][estimated a $25\%$ contribution in the V band due to the disc]{chevalier89}, while \cite[][]{shahbaz93} estimated a 10\% contribution in the IR. Concerning the UVW1 band, the maximum count rate is 0.7 c/s, this corresponds to an increase by a factor of $\sim34$ compared to the ground level.  This implies that the intrinsic emission in this band must be lower than $3\%$.

Following our analysis for the X-ray and UV fluxes, we can also measure the ratio between X-ray and optical fluxes. However, we only consider here the variable optical flux ($3\times10^{-13}$ \ergscm, about 60\% of the total). Taking into account the uncertainty on the X-ray flux measurement the resulting ratio is $0.6\%<F_{O}/F_{X}<6.5\%$. We conclude that also the variable optical emission is likely due to reprocessing from the companion star.

Summarizing, we conclude that the accretion disc must be intrinsically faint in the UV during quiescence, with a maximum of $3\%$ of the total light in this band arising from it, while, it can be brighter in the optical band. This is in agreement with the conclusion of \cite{torres02} that suggested that the companion has a comparable luminosity in the optical (H$_{\alpha}$) than the accretion disc and is brighter at shorter wavelengths. 
This is also in agreement with the analysis of \cite{reynolds08} that with the goal of understanding the nature of a NIR excess in the emission from a sample of X-ray Novae, hosting both BH and NS, including Cen X-4, found that 
the component required to match the observed excess is consistent with a
disc truncated at $10^{3}-10^{5} R_{s}$, with an irradiated temperature profile in all cases
(i.e., $T(r)\sim r^{-0.5}$). A standard steady state temperature profile
is unable to account for the observed excess. Such kind of disc will contribute only
minimally in the UV.
We also conclude that almost all the variable UV emission, which represents $\sim100\%$ of the total emission in this bands, must arise from reprocessing from the surface of the companion facing the NS.  Also part of the variable optical (V) emission, that represents $\sim60\%$ of the total in the band, must arise from reprocessing, while another part can arise from orbital modulation. 
By folding the V and UVW1 light curves at the 15.1 orbital period, we measured a 6\% and 10\% $3\sigma$ upper limit due to the orbital contribution respectively \citep[however,][with higher S/N data, measured the V flux variation due to orbital 15.1 h modulation to be $\sim17\%$]{chevalier89}. Moreover, we also found that the X-ray irradiation is likely generating on the surface of the companion a small ($\sim2\%$ of the whole surface for a $0.5R_{\odot}$ companion) hot spot. 

Finally, we emphasize that the ADAF and reprocessing scenario are not mutually exclusive, they in fact could easily coexist. Indeed, if most if not all the UV emission is produced by reprocessing from the companion star, as we have shown, it is not surprising that the ADAF model alone can not account for the intense emission in the UV band \citep[as suggested by][]{menou01}.  We note that, at least for what concerns Cen X-4, a detailed model including ADAF plus reprocessing from the companion should be tested against the data. We have demonstrated that matter must accrete on the NS surface, and the fact that this matter can be in an ADAF state is not in contrast with our results. 

We emphasize that the estimate of the ratio $F_{UV}/F_{X}$ depends on the exact slope of the hard power law tail. New simultaneous observations performed e.g. with {\it NuSTAR} and {\it Swift} (or {\it XMM-Newton} or Hubble Space Telescope, {\it HST}) will certainly allow to firmly measure the slope and the flux of the power law ({\it NuSTAR}) together with the simultaneous flux in the optical and UV bands ({\it Swift} or {\it HST} or {\it XMM-Newton}). This will constrain 
the ratio $F_{UV}/F_{X}$. If this ratio will be well above $4.7\%$ \footnote{Considering the $3\sigma$ uncertainty on the mass ratio $q\lesssim0.29$, consequently, $F_{UV}/F_{X}\lesssim3.9\%$. By including also the maximum contribution from the accretion disc 0.8\% we get a total of about 4.7\%.}, 
a different scenario from reprocessing should be considered. 
Indeed, no other optically thick surfaces are present in the binary system other than the disc and the companion star. The only plausible alternative to it, is that the accreting matter is intrinsically emitting in the UV. The specific ADAF model for this system should be produced accordingly.

\section{Summary}
\label{sec:summary}

We analyzed 60 multiwavelength (O, UV and X-ray) simultaneous observations of Cen X-4, performed by the \Sw\ observatory on a daily basis, with the goal to learn from the source quiescent variability about the emission mechanisms powering the optical, UV and X-ray emission. The main result of this work can be summarized as follow:

The optical, UV, and X-ray quiescent light curves are strongly variable both on short (days) than on long (months) timescale as showed by their first order structure function. The X-ray band is more variable than the UV and the optical band, meaning that the amplitude of the variability is higher, as also pointed out by the fractional root mean square variability which is F$^{X-ray}_{\rm var}=73.0\pm1.5\%$, $F^{UVW1}_{\rm var}=50.0\pm1.4\%$, $F^{V}_{\rm var}=10.0\pm1.6\%$. We observed the strongest short timescale X-ray, flare-like, variability ever recorded from Cen X-4 in quiescence: a factor of 22 drop in only 4 days.

We found a highly significant and strong correlation between the X-ray and the UV (UVW2, UVM2, UVW1, U) emission on long timescale (t$\leq5$ ks), on short timescale (t$\leq2$ ks), and also on very short timescale (down to t$\leq$110 s).  We found a significant but less intense correlation also between the X-ray and the V band on long timescale (t$\leq5$ ks), and on short timescale (t$\leq2$ ks). No significant correlation was detected with the B band, likely because the limited number of pointings performed in this band and the consequent lack of coverage at high X-ray count rate. The most likely shape of the correlation is a power law with index $\Gamma=0.2-0.6$. 
 
We produced three averaged X-ray spectra as a function of the source count rate, low $<0.07$ c/s, medium 0.07--0.11 c/s, and high $>0.11$ c/s. We found the spectra well fitted by a model made by the sum of a thermal component  in the form of hydrogen NS atmosphere plus a non-thermal power law tail, both multiplied by the interstellar absorption. Both components must change with the flux. The overall spectral shape remains the same with respect to flux variability as we show by both count rate and spectral analysis. The thermal and the power law component are changing in tandem, being both responsible always for the same fraction, about 50\%, of the total X-ray flux in the 0.5--10 keV band. This suggests that they are physically linked. The temperature of the NS surface is changing being hotter when the flux is higher: kT$_{h}=80\pm2$ eV, kT$_{m}=73.4\pm0.9$ eV, kT$_{l}=59.0\pm1.5$ eV (while $\Gamma_{h}=1.4\pm0.5$, $\Gamma_{m}=1.4\pm0.2$, $\Gamma_{l}=2.0\pm0.2$). We conclude that accretion is very likely still occurring also at low Eddington luminosity rates ($\sim10^{-6}$) and that the matter is reaching and heating the NS surface, overtaking the action of the centrifugal force of the rotating magnetosphere. Since the two spectral component are closely linked, accretion is likely generating somehow also the power law component.
  
We demonstrated that the UV emission cannot be thermal emission from the gas stream-impact point. This scenario can not account for the timescale of the observed X-ray and UV correlation. The UV emission can not arise neither from matter moving inward in a standard optically thick and geometrically thin accretion disc. Indeed, due to the short timescale of the correlation the matter should move from a starting point at a maximum distance of about 30 km from the star center. However, the disc is very likely not extending so close to the surface, otherwise it would shine in the X-ray, moreover, it has never been observed in the X-ray quiescent spectrum of Cen X-4. The UV emission is not even produced by reprocessing from a standard thin disc. Indeed, the fraction of UV vs X-ray flux is too high (in the range $2.1\%-23\%$), as the maximum allowed reprocessing fraction from a standard disc is lower then $0.8\%$. We found, on the contrary, that most (if not all) the UV emission and part ($\sim60\%$) of the optical emission is very likely generated by reprocessing from the whole surface of the companion star. Moreover, we found indication of the presence of a small hot spot  ($\sim16000$ K, $\sim2.4\times10^{4}$ km, fractional area of about $2\%$) on the companion star which could be generated by X-ray irradiation, likely in a region where no shield is provided by the accretion disc. We found that the accretion disc must be UV faint, as its contribution to the total UV light is less than $3\%$ (but it can still be optically bright).

We have shown that a strong propeller-type outflow powered by the interaction with the accretion flow and the stellar magnetic field is very unlikely to be present in Cen X-4, as is required for the ADAF-propeller scenario.
A strong outflowing wind (ADIOS) solution could be instead favoured. However, a `dead disk' may also form in the inner regions of the accretion flow, which inhibits but does not completely prevent accretion onto the star, allowing accretion to proceed even at rates when the 'propeller' solution would otherwise be active.

The timescale of the correlation could suggests that the accreting matter can be in an ADAF state, which is likely UV faint, at a distance less or equal to $\sim6200\,R_{s}$. However, the spectral analysis (SED) does not confirm the presence of an ADAF. We have only tested separate reprocessing and ADAF models, however, it could be that both processes are involved. A more detailed model including both an ADAF and reprocessing on the companion star should be developed and tested against these data.

\section*{Acknowledgments}
FB thanks Aron Zell and Jean Charles Tropato for their precious help, and Ramesh Narayan for the useful discussion. ND is supported by NASA through Hubble Postdoctoral Fellowship grant number HST-HF-51287.01-A from the Space Telescope Science Institute. RW acknowledges support by an European Research Council Starting Grant.

\bibliographystyle{mn2e}
\bibliography{biblio}

\vfill\eject
\end{document}